\DocumentMetadata{}
\documentclass[sigconf,balance=false]{acmart}
\usepackage{popets}
\setcopyright{popets}
\copyrightyear{YYYY}
\acmYear{YYYY}
\acmVolume{YYYY}
\acmNumber{X}
\acmDOI{XXXXXXX.XXXXXXX}
\acmISBN{}
\acmConference{Proceedings on Privacy Enhancing Technologies}
\renewcommand\footnotetextcopyrightpermission[1]{}
\usepackage{soul}
\usepackage{tikz}
\usepackage{colortbl}
\usepackage{cellspace}
\PassOptionsToPackage{x11names}{xcolor}
\usetikzlibrary{arrows, patterns, automata, arrows.meta, positioning, shapes.geometric, backgrounds, fit, calc}

\usepackage{amssymb}
\usepackage{amsfonts}
\usepackage{textcomp}
\usepackage{xcolor}
\def\BibTeX{{\rm B\kern-.05em{\sc i\kern-.025em b}\kern-.08em
    T\kern-.1667em\lower.7ex\hbox{E}\kern-.125emX}}
\usepackage[utf8]{inputenc}
\usepackage{graphicx}
\usepackage{subfigure}
\usepackage{units}
\usepackage{accents}
\usepackage{blindtext}
\usepackage{multirow}
\usepackage{xspace}
\usepackage{amsmath}
\usepackage{paralist}
\usepackage{mdframed}
\usepackage{pict2e}
\usepackage{listings}
\usepackage{color}
\usepackage{threeparttable}
\usepackage{array}
\usepackage{booktabs}
\usepackage{pifont}
\usepackage{balance}
\usepackage{url}

\usepackage{tabularx}
\usepackage{caption}
\usepackage{lipsum}
\usepackage{makecell}
\usepackage[most]{tcolorbox}
\usepackage{amsbsy}
\usepackage{hhline}
\usepackage{mathtools}

\usepackage[symbol]{footmisc}
\renewcommand{\thefootnote}{\fnsymbol{footnote}}

\usepackage[ruled]{algorithm2e}
\usepackage{algorithmic}
\LinesNumbered

\usepackage{url}

\usepackage[
    n,
    advantage,
    operators,
    sets,
    adversary,
    landau,
    probability,
    notions,
    logic,
    ff,
    mm,
    primitives,
    events,
    complexity,
    asymptotics,
    keys]{cryptocode}

\definecolor{burgundy}{RGB}{144,0,32}
\definecolor{myblue}{rgb}{0.0, 0.2, 0.8}
\definecolor{bluegray}{rgb}{0.4, 0.6, 0.8}
\definecolor{azure}{rgb}{0.0, 0.5, 1.0}
\definecolor{darkcandyapplered}{rgb}{0.64, 0.0, 0.0}
\definecolor{blue-violet}{rgb}{0.54, 0.17, 0.89}
\definecolor{applegreen}{rgb}{0.55, 0.71, 0.0}

\newcolumntype{?}{!{\vrule width 1pt}}
\newlength{\arrayrulewidthOriginal}

\newcommand{\dbpaisalong}{Discovery-Based Privacy-Agile IoT Sensing+Actuation}
\newcommand{\paisa}{{{\sf PAISA}}\xspace}
\newcommand{\dbpaisa}{{{\sf DB-PAISA}}\xspace}
\newcommand{\dbb}{{{\sf DB-}}\xspace}
\newcommand{\paisaplus}{\dbpaisa}

\newcommand{\dev}{\ensuremath{dev}\xspace}

\newcommand{\iot}{\ensuremath{IoT}\xspace}
\newcommand{\usr}{\ensuremath{usr}\xspace}
\newcommand{\owner}{\ensuremath{owner}\xspace}
\newcommand{\iiot}{\ensuremath{IIoT}\xspace}
\newcommand{\iotdev}{\ensuremath{{\mathsf I_{\mathsf \dev}}}\xspace}
\newcommand{\iotdevpk}{\ensuremath{pk_{\mathsf \iot}\xspace}}
\newcommand{\iotdevsk}{\ensuremath{sk_{\mathsf \iot}\xspace}}
\newcommand{\usrdev}{\ensuremath{{\mathsf U_{\mathsf \dev}}}\xspace}
\newcommand{\ownerdev}{\ensuremath{{\mathsf O}_{\dev}}\xspace}
\newcommand{\ownerpk}{\ensuremath{pk_{\mathsf \owner}\xspace}}
\newcommand{\ownersk}{\ensuremath{sk_{\mathsf \owner}\xspace}}
\newcommand{\mfr}{{\ensuremath{\mathsf M}{fr}}\xspace}
\newcommand{\mfrpk}{\ensuremath{pk_{\mathsf \mfr}\xspace}}

\newcommand{\iotdevsoft}{\ensuremath{SW_{\dev}\xspace}}

\newcommand{\iotdevsofthash}{\ensuremath{H_{\iotdevsoft}\xspace}}

\newcommand{\pubtcb}{\ensuremath{SW_{\pub}\xspace}}
\newcommand{\privatetcb}{\ensuremath{SW_{\private}\xspace}}
\newcommand{\pubid}{\ensuremath{ID_{\pub}\xspace}}

\newcommand{\rot}{\texttt{RoT}\xspace}

\newcommand{\seckey}{\ensuremath{\mathcal K}\xspace}
\newcommand{\certiot}{\ensuremath{\mathsf{Cert_{\iot}}}\xspace}
\newcommand{\certmfr}{\ensuremath{\mathsf{Cert_{\mfr}}}\xspace}
\newcommand{\devmanifest}{\ensuremath{\mathsf{Manifest_{\mathsf \iot}}}\xspace}

\newcommand{\devmanifesturl}{\ensuremath{\mathsf{URL_{Man}}}\xspace}
\newcommand{\devmanifesturlfull}{\ensuremath{\mathsf{URL_{Man_{Full}}}}\xspace}

\newcommand{\sigfunc}{\ensuremath{{\ensuremath{\sf{\mathcal SIG}}}\xspace}}
\newcommand{\sigvalue}[1]{\ensuremath{\mathsf{Sig_{#1}}}\xspace}

\newcommand{\noncevalue}[2]{\ensuremath{\mathsf{N_{#1}^{#2}}}\xspace}
\newcommand{\tsvalue}[1]{\ensuremath{\mathsf{time_{#1}}}\xspace}
\newcommand{\busytime}{\ensuremath{\mathsf{U_{Busy}}}\xspace}

\newcommand{\push}{\ensuremath{Push}\xspace}
\newcommand{\pull}{\ensuremath{Pull}\xspace}

\newcommand{\pub}{\dbpaisa}
\newcommand{\private}{{{\sf IM-PAISA}}\xspace}

\newcommand{\wait}{{\ensuremath{\sf{\mathcal Wait}}}\xspace}
\newcommand{\nlist}{\ensuremath{\mathsf{Pool_{N}}}\xspace}
\newcommand{\ntmplist}{\ensuremath{\mathsf{Pool_{{N}_{Tmp}}}}\xspace}
\newcommand{\poolsize}{\ensuremath{\mathsf{|Pool_{N}|_{Max}}}\xspace}
\newcommand{\ntmplistsize}{\ensuremath{\mathsf{|Pool_{{N}_{Tmp}}|}}\xspace}
\newcommand{\gen}{{\ensuremath{\sf{\mathcal Gen}}}\xspace}
\newcommand{\rcv}{{\ensuremath{\sf{\mathcal Rcv}}}\xspace}

\newcommand{\geninterval}{{\ensuremath{\sf{T_{\gen}}}}\xspace}
\newcommand{\ann}{{\ensuremath{\sf{\mathcal Ann}}}\xspace}
\newcommand{\anninterval}{{\ensuremath{\sf{T_{\ann}}}}\xspace}

\newcommand{\req}{\ensuremath{req}\xspace}
\newcommand{\resp}{\ensuremath{resp}\xspace}
\newcommand{\responsemessage}{{\ensuremath{\sf{Msg_{\resp}}}}\xspace}
\newcommand{\requestmessage}{{\ensuremath{\sf{Msg_{\req}}}}\xspace}
\newcommand{\attest}{{\ensuremath{\sf{\mathcal Att}}}\xspace}
\newcommand{\attestresult}{\ensuremath{\mathsf{Att_{result}}}\xspace}
\newcommand{\attestreport}{\ensuremath{\mathsf{Att_{report}}}\xspace}
\newcommand{\attestinterval}{{\ensuremath{\sf{T_{\attest}}}}\xspace}

\newcommand{\verifystep}{{\ensuremath{\sf{\mathcal Verify}}}\xspace}

\newcommand{\registration}{{\ensuremath{\mathsf{\it Registration}}}\xspace}

\newcommand{\runtime}{{\ensuremath{\mathsf{\it Runtime}}}\xspace}

\newcommand{\response}{\ensuremath{\mathsf{\sf Response}}\xspace}
\newcommand{\requestinterval}{{\ensuremath{\sf{T_{\mathcal Req}}}}\xspace}

\newcommand{\request}{\ensuremath{\mathsf{\sf Request}}\xspace}

\newcommand{\receptionusr}{\ensuremath{\mathsf{\sf Reception}}\xspace}

\newcommand{\prv}{{\ensuremath{\sf{\mathcal Prv}}}\xspace}
\newcommand{\vrf}{{\ensuremath{\sf{\mathcal Vrf}}}\xspace}
\newcommand{\ra}{{\ensuremath{\sf{\mathcal RA}}}\xspace}
\newcommand{\sadv}{{\ensuremath{\sf{\mathcal Adv}}}\xspace}
\newcommand{\chal}{{\ensuremath{\sf{\mathcal Chal}}}\xspace}

\def\thickhline{\noalign{\hrule height1pt}}
\mathchardef\mhyphen="2D

\begin{document}

\title[\dbpaisa]{\dbpaisa: \dbpaisalong}

\author{Isita Bagayatkar}
\email{ibagayat@uci.edu}
\affiliation{%
   \institution{University of California, Irvine}
   \city{}
   \country{}
}

\author{Youngil Kim}
\email{youngik2@uci.edu}
\affiliation{%
   \institution{University of California, Irvine}
   \city{}
   \country{}
}

\author{Gene Tsudik}
\email{gene.tsudik@uci.edu}
\affiliation{%
   \institution{University of California, Irvine}
   \city{}
   \country{}
}

\renewcommand{\shortauthors}{Isita Bagayatkar, Youngil Kim, \& Gene Tsudik}

\begin{abstract}
Internet of Things (IoT) devices are becoming increasingly commonplace in numerous public and semi-private settings. 
Currently, most such devices lack mechanisms to facilitate their discovery by casual (nearby) users who are not
owners or operators. However, these users are potentially being sensed, and/or actuated upon, by these devices, 
without their knowledge or consent. This naturally triggers privacy, security, and safety issues.

To address this problem, some recent work explored device transparency in the IoT ecosystem. The intuitive approach 
is for each device to periodically and securely broadcast (announce) its presence and capabilities to all nearby users. 
While effective, when no new users are present, this \push-based approach generates a substantial amount of 
unnecessary network traffic and needlessly interferes with normal device operation.

In this work, we construct \dbpaisa\ which addresses these issues via a \pull-based method, whereby devices 
reveal their presence and capabilities only upon explicit user request. Each device 
guarantees a secure timely response (even if fully compromised by malware) based on a small 
active Root-of-Trust (RoT). \dbpaisa\ requires no hardware modifications and is suitable for a  
range of current IoT devices. To demonstrate its feasibility and practicality, we built a fully functional and 
publicly available prototype. It is implemented atop a commodity MCU (NXP LCP55S69) and operates in tandem with a 
smartphone-based app. Using this prototype, we evaluate energy consumption and other performance factors.
\end{abstract}
\renewcommand{\thefootnote}{\arabic{footnote}}

\maketitle

\section{Introduction \label{sec:intro}}
In recent years, Internet of Things (IoT), embedded, and smart 
devices have become commonplace in many aspects of everyday life. 
We are often surrounded by cameras, sensors, displays, and robotic appliances 
in our homes, offices, public spaces, and industrial settings.
With rapid advances in AI/ML, 5G, robotics, and automation,
the use of IoT devices is becoming more and more prevalent.
 
IoT devices feature various sensors and/or actuators. 
Sensors collect information about the environment, while actuators control the environment.
Some IoT devices use sensors to collect sensitive information, e.g., cameras, voice assistants, and 
motion detectors. Whereas, actuators on some IoT devices perform safety-critical tasks, e.g., operate door locks, 
set off alarms, and control smart appliances. This is not
generally viewed as a problem for owners who install, deploy, and operate these devices. 
After all, owners should be aware of their devices' locations and functionalities.
However, the same does not hold for casual users who come within sensing and/or 
actuation range of IoT devices owned and operated by others.
Such users remain unaware unless nearby devices are human-perceivable, e.g., visible or audible.

This issue is partly due to lack, or inadequacy, of security features on commodity IoT devices,
which manufacturers often justify with size, energy, and monetary constraints of the devices.
Moreover, consumers gravitate towards monocultures, as witnessed by the great popularity 
of certain devices, such as Ring doorbells, Roomba vacuum cleaners, and Echo voice assistants.
This leads to such massively popular IoT devices becoming highly attractive attack targets.
Attacks can exploit devices' software vulnerabilities to exfiltrate 
sensed data, report fake sensed data, perform malicious actuation, or simply zombify devices
\cite{mirai,miraianalysis,stuxnet,miller2015remote,butun2019security,sikder2021,smarthome-traffic-snoop,home-sec-sok}. 
To mitigate these risks, a large body of work focused on constructing small Roots-of-Trust (\rot-s) for 
IoT devices. This includes remote attestation (\ra) 
\cite{smart,vrasedp,sanctum,tan2011tpm,pioneer}, run-time 
attestation \cite{tinycfa,litehax,cflat,caulfield2023acfa}, and sensor data protection 
\cite{sancus,pfb,fernandes2016flowfence}. However, all these techniques are research proposals;
they are largely ignored by manufacturers who lack compelling incentives to introduce security features to their products.

Furthermore, most prior work on privacy and security for IoT ecosystems 
\cite{awdt-dominance,garota,casu,song2017privacy,dwivedi2019decentralized} and a few
government IoT guidelines  \cite{nist-reco,aus-code-practice,uk-code-practice} focus on
device owners or operators who are well aware of device presence and capabilities.
As mentioned above, IoT devices also impact other users within the sensing and/or actuation range. Ideally, 
these casual nearby users must be informed of the device's presence and capabilities,
which would enable users to make informed decisions.

Another motivation comes from data protection laws, such as the European General Data 
Protection Regulation (GDPR) \cite{gdpr} and California Consumer Privacy Act (CCPA) \cite{ccpa}. 
They aim to protect individuals' personal data and grant them control over data collection, processing, 
storing, and sharing. We observe that many (perhaps most) IoT devices operate in tandem
with a cloud-based {\em digital twin} hosted by device manufacturers, software vendors, or users.
In such cases, sensed data is processed and stored on devices as well as in the cloud.
Therefore, logic dictates that IoT devices should also be subject to the same data protection laws. 

Privacy and safety risks are not only apparent in public spaces, e.g., city streets,
event venues, parks, campgrounds, airports, and beaches. They also occur in semi-private 
settings, e.g., conference rooms, hotels, and private rentals, such as Airbnb. In such places, 
privacy is expected, yet not guaranteed \cite{yao2019defending,tenant-hidden-camera,airbnb-news}. 
Casual visitors (users) are naturally suspicious of unfamiliar surroundings \cite{inman, enduser-privacy} 
partly due to being unaware of nearby IoT devices.

\subsection{Current State of Device Transparency} \label{subsec:dev-transparency}
To address the issues presented above, some recent research explored a privacy-agile IoT 
ecosystem based on manufacturer compliance. In particular, \paisa \cite{paisa} allows
users to learn about the presence and capabilities of nearby devices by listening to periodic secure
announcements by each device using WiFi broadcast. Announcements are guaranteed
to be sent in time even if the device is fully malware-compromised. Also, each announcement includes
a recent attestation token, allowing a recipient user to check whether the device is trustworthy.
\paisa works on commodity devices that have a Trusted Execution Environment (TEE) (e.g., ARM 
TrustZone \cite{trustzone}). It is unsuitable for low-end IoT devices without TEE support.

Another recent work, DIAL \cite{dial}, requires each device to have a physically attached NFC tag.
The device helps users locate it by either sounding a buzzer or using 
an ultra-wideband (UWB) interface.  Upon physically locating a device, 
the user simply taps a smartphone on the NFC tag to get device information. Although DIAL does not 
require a TEE on each device, it imposes other requirements on the manufacturer and user, 
such as mounting an NFC tag and a buzzer or a UWB interface (rare on commodity IoT devices) 
for localization. It also requires physical access to a device to tap the NFC tag manually.

Both \paisa\ and DIAL follow the \push model: IoT devices announce their presence at 
fixed time intervals. This raises two concerns: (1) {\it unnecessary network traffic}, 
the volume of which can be high when numerous compliant devices are deployed in a given
space, and (2) {\it interference with normal device operation} (in \paisa only), which is 
detrimental for devices that perform safety- or time-critical tasks.
Both (1) and (2) are especially problematic when no new users are around to scan for 
these announcements. Also, as discussed above, DIAL's requirements for NFC-s and buzzers or
UWB interfaces are currently unrealistic for most settings.

Another common application domain is large IoT deployments in 
industrial settings. Industrial IoT (\iiot) devices play an increasingly important role in 
automation across various industry sectors. Their numbers surpassed $3.5$ billion in 2023, 
accounting for $25$\% of total IoT deployments \cite {statista-iiot,statista-iot}.
According to a recent Ericsson mobility report \cite{ericsson-report}, the estimated 
number of connected devices in a typical smart factory is between 0.5 and 1 per square meter.
This implies potentially millions of devices in a large industrial installation, 
e.g., a warehouse, port, or factory \cite{large-factories}. 

In such large \iiot\ systems, the owner/operator needs to maintain and keep track of all deployed devices.
Unlike public and semi-public settings with casual users, the owner knows the identities and types of
deployed \iiot\ devices. However, they might not know which devices are reachable, operational, or 
malware-compromised. They might also not know device locations since many industrial settings involve 
moving components. An intuitive approach is to use existing infrastructure to probe devices.
For example, specialized software for managing merchandise or IoT devices associated with 
tagging information (e.g., barcode or NFC) can be used for inventory management 
\cite{tejesh2018warehouse,kobbacy1999towards,atieh2016performance,zahran2021iiot}. However, these techniques 
do not provide either reliable {\it current software state} \  or {\it current location} of moving devices. 

\subsection{Overview and Contributions}
In this paper, we construct (1) \paisaplus, an efficient privacy-compliant IoT technique geared
for (semi-)public environments, and (2) \private: {\it \underline{I}nventory \underline{M}anagement-PAISA}, geared
for \iiot settings, described in Appendix \ref{sec:inventory}.
Both techniques are \pull-based, meaning that a user (or an owner in \private) 
issues an explicit request to learn about nearby devices. This obviates the need to generate 
and broadcast device announcements continuously. 
As a consequence, it reduces network load and interference 
with normal device functionality, especially when no new users are nearby.
This additional communication step of sending a discovery request changes the adversary 
model, which results in new security challenges, discussed later in the paper.

\paisaplus has two primary components: (1) a user device that sends a request and processes responses, and (2) compliant IoT devices that ensure 
a response containing software status and capabilities upon request.
A manufacturer specifies the device information that can be released by its IoT devices.

Unlike relevant prior work \cite{garota,casu,pfb}, \paisaplus requires no hardware modifications 
for low-end devices and relies on a popular TEE (ARM TrustZone \cite{ARM-TrustZone-M})
to prioritize its tasks  over other software. 
Also, \paisaplus imposes no special requirements on user devices,
except that IoT and user devices should support the same network interface, e.g., WiFi or Bluetooth.

Contributions of this work are three-fold:
\begin{compactitem}
    \item \pub, a pull-based IoT device discovery framework with low bandwidth overhead and 
    no interference with normal device operation whenever no new users are present.
    \item A comparison of bandwidth, energy, and runtime overheads between \paisa's \push and \paisaplus's \pull models. 
    \item A full prototype of \pub comprised of: (a) an IoT device with ARM TrustZone-M and 
    Bluetooth extended advertisements, and (b) an Android app that requests, scans, processes, 
    and displays IoT device information. The implementation is publicly available at \cite{paisaplus-code}.
\end{compactitem}

\section{Background}\label{sec:background}

\begin{figure}[t]
  \centering  
  \captionsetup{justification=centering}  
  \includegraphics[width=\columnwidth]{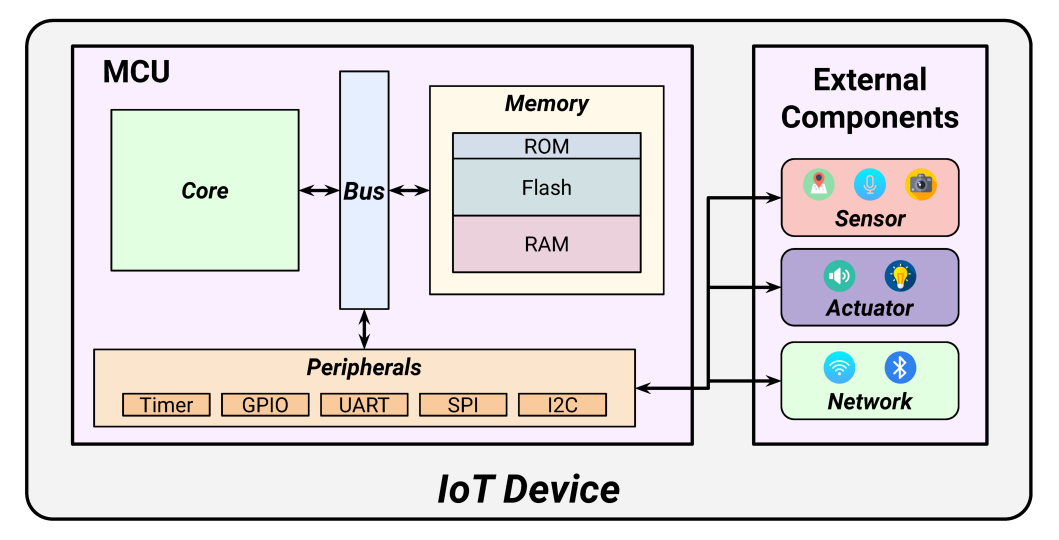}
  \vspace{-.6cm}
  \caption{IoT Device Layout}
  \label{fig:device_layout}
  \vspace{-.2cm}
\end{figure}

\subsection{Targeted IoT Devices}\label{subsec:bg_target}
In general, we consider constrained IoT devices geared for executing simple tasks and 
deployed on a large scale, e.g., smart bulbs, locks, speakers, plugs, and alarms. 
Given tight resource budgets, they typically rely on microcontroller units (MCUs) with 
no memory virtualization.
Many of such MCUs have ARM Cortex-M \cite{cortexM} or RISC-V cores \cite{risc-v}.
Our focus is on devices equipped with relatively simple TEEs, such as TrustZone-M.
Note that we explicitly rule out low-end MCUs with no hardware security features.

As shown in Figure \ref{fig:device_layout}, a typical IoT device contains an MCU and multiple 
peripherals. The MCU includes a core processor(s), main memory, and memory bus, forming a 
System-on-a-Chip (SoC).
Its primary memory is typically partitioned into: (1) program memory (e.g., flash), 
where software is installed, (2) data memory (RAM), which the software uses for its stack 
and heap, and (3) read-only memory (ROM), where the bootloader and tiny immutable 
software are placed during provisioning.
Such an MCU also hosts peripherals, including a timer, General-Purpose Input/Output (GPIO), and
Universal Asynchronous Receiver/Transmitter (UART).

The MCU interfaces with various special-purpose sensors and actuators through such peripherals.
Common sensors are exemplified by microphones, GPS units, cameras, gyroscopes, as well as
touch and motion detectors. Whereas, speakers, door locks, buzzers, sprinklers, 
as well as light and power switches, are examples of actuators.

{\noindent \bf Scope of Targeted IoT Device Types:} For certain personal IoT devices, 
there is no need to inform nearby users of their presence since doing so can 
compromise their owners' privacy. This involves medical devices, e.g., blood pressure monitors,
insulin pumps, catheters, or pacemakers. Clearly, neither \paisa\ nor \paisaplus\ is 
applicable to such devices. Delineating the exact boundaries between devices' 
presence of which should (or should not) be released (or be discoverable) is 
beyond the scope of this paper.

\subsection{Network}\label{subsec:bg_network}
IoT devices are connected to the Internet and/or other devices either directly or through 
intermediate entities, e.g., hubs or routers. To support this, a typical IoT device features 
at least one network interface (e.g., WiFi, Bluetooth, Cellular, Ethernet, or Zigbee).
WiFi and Cellular provide wireless Internet connectivity at relatively high speeds, 
while Bluetooth and Zigbee are geared towards lower-speed, shorter-range communication 
with neighboring devices. In this work, we focus on WiFi and Bluetooth since they 
are commonplace on both smartphones and IoT devices \cite{blevswifi}.

{\noindent \bf WiFi} 802.11n (aka WiFi 4) \cite{wifi-802-11-n} has a range of $\leq~75$m 
indoors and $\leq~250$m outdoors \cite{abdelrahman2015comparison}.
The latest WiFi standards (e.g., WiFi 6) achieve multi-gigabit speeds, making it ideal for 
bandwidth-intensive activities.

{\noindent \bf Bluetooth} 5 \cite{bluetooth-5} operates on a relatively shorter range, 
typically $\leq~40$ meters indoors \cite{collotta2018bluetooth} and its data transfer 
speed peaks at $\approx~2$Mbps.
Since Bluetooth 4.0, Bluetooth devices have started supporting Bluetooth Low Energy (BLE).
BLE is optimized for low power consumption, making it well-suited for resource-constrained IoT devices. 
Bluetooth Classic is utilized for connection-oriented data transfer, such as audio streaming.
Whereas, BLE is widely used in beaconing applications (e.g., asset tracking, proximity marketing, and indoor navigation) 
for its power efficiency.

\subsection{Trusted Execution Environment (TEE) \label{subsec:bg_tee}}
A TEE is a secure and isolated environment within a computer system, 
typically implemented as a hardware-based component. It provides a secure area for the 
execution of sensitive operations (e.g., cryptographic processing or handling of sensitive data) 
without interference or compromise from other software, operating systems, and hypervisors.

ARM TrustZone, one of the most prominent commercial TEE-s, divides the system into two separate 
execution environments: Secure world and Normal (non-secure) world. Non-secure applications and interrupts 
cannot access computing resources, such as memory and peripherals, that are configured as secure.
Within the ARM TrustZone framework, TrustZone-A (TZ-A) and TrustZone-M (TZ-M) refer to two specific 
implementations tailored for different types of processors; TZ-A is designed for high-performance 
application processors (e.g., smartphones), while TZ-M is tailored for low-power, resource-constrained 
MCUs commonly used in IoT devices. Although core security goals are the same for both TZ-A and TZ-M, 
techniques used to achieve these goals differ.

Unlike TZ-A, TZ-M is memory map-based.
Memory areas and other critical resources marked as secure can only be accessed when the core is 
running in a secure state. Peripherals can also be designated as {\it secure},
which makes them exclusively accessible through and controlled by secure code.
Besides two security regions (Secure world/Normal world), TZ-M introduces a non-secure callable (NSC) region.
The NSC region allows secure functions to be safely called from non-secure code, 
exposing certain secure functions to Normal world without compromising overall system security.

\begin{figure*}
	\centering
	\subfigure[Interactive (Traditional) \ra \label{fig:interactive_ra}]
	{\includegraphics[width=\columnwidth]{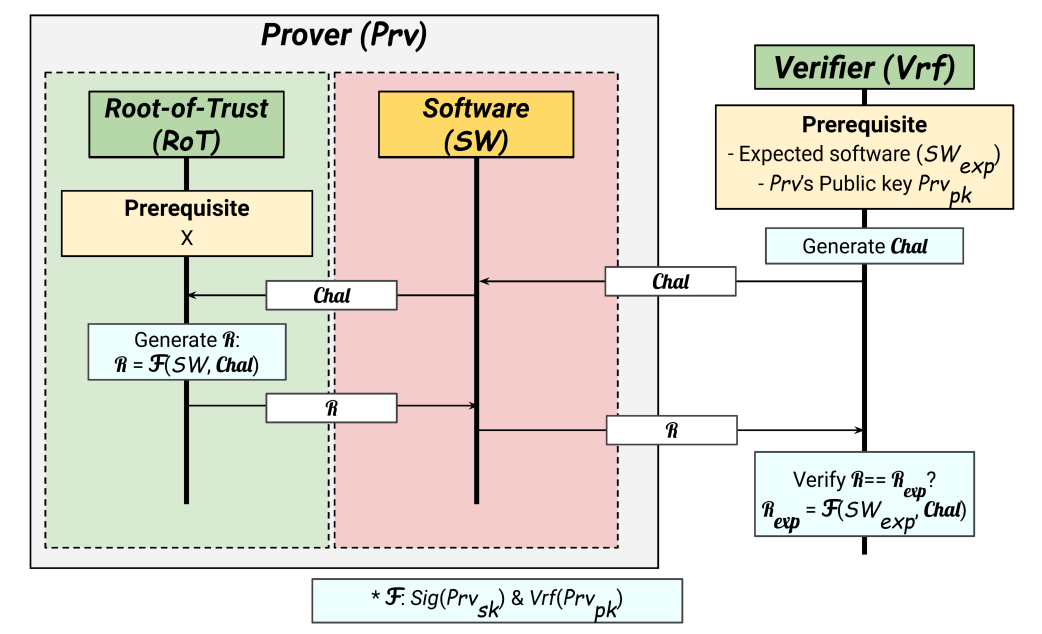}}    
	\subfigure[Non-Interactive and Self-Initiated \ra \label{fig:non_interactive_ra}]
	{\includegraphics[width=\columnwidth]{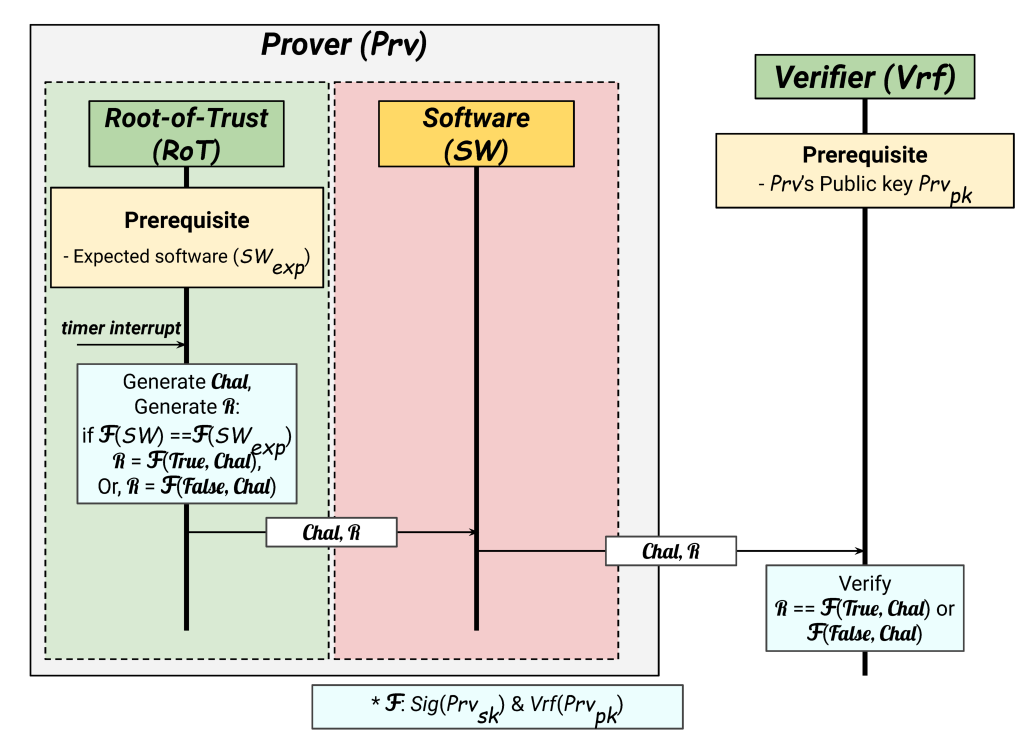}}
      \vspace*{-0.3cm}
	\caption{Different Types of \ra}	
     \vspace*{-0.2cm}
	\label{fig:ra} 
\end{figure*}

\subsection{Remote Attestation (\ra)}\label{subsec:bg_ra}
\ra is a security technique for verifying integrity and trustworthiness of software
state on a remote entity. It allows a trusted server (verifier -- \vrf) to securely 
determine whether a remote device (prover -- \prv) is running the expected software. 
As shown in 
Figure \ref{fig:interactive_ra}, \ra is typically realized 
as a challenge-response protocol:
\begin{compactenum}  
    \item \vrf initiates \ra by sending a request with a unique challenge 
    (\chal) to \prv.
    \item \prv generates an unforgeable attestation report, which includes an authenticated 
    integrity check over its software and \chal, and returns it to \vrf.
    \item \vrf verifies the attestation report and determines \prv's current state.
\end{compactenum}
Figure \ref{fig:non_interactive_ra} shows a variation of the above protocol, referred to as non-interactive \ra \cite{ambrosin2020collective}.
In the variant, \prv autonomously decides when to trigger \ra and locally generates a unique
and timely \ra report with \chal. This obviates the need for \vrf to initiate the \ra process.
Also, since \prv knows the (hash of) benign software that it is supposed to 
execute, it can generate its own \ra reports indicating whether or not it is indeed running the expected software.
This way, \vrf no longer needs to know the expected software configuration for each \prv.

\section{Design Overview} \label{sec:overview}
\subsection{\pub Protocol Overview}
\pub involves two main entities: an IoT device (\iotdev) and a user device (\usrdev).
\iotdev is equipped with a TEE and deployed in public or semi-private settings.
\usrdev can be any personal device with sufficient computing resources, such as a smartphone,
smartwatch, or AR device. 
Also, a device manufacturer (\mfr) serves a minor (yet trusted)
role in creating and maintaining accurate and up-to-date information for its \iotdev-s.

\begin{figure*}
        \vspace*{-0.3cm}
	\centering
	\subfigure[\paisa \push Model Overview \label{fig:push}]
	{\includegraphics[width=0.49\textwidth]{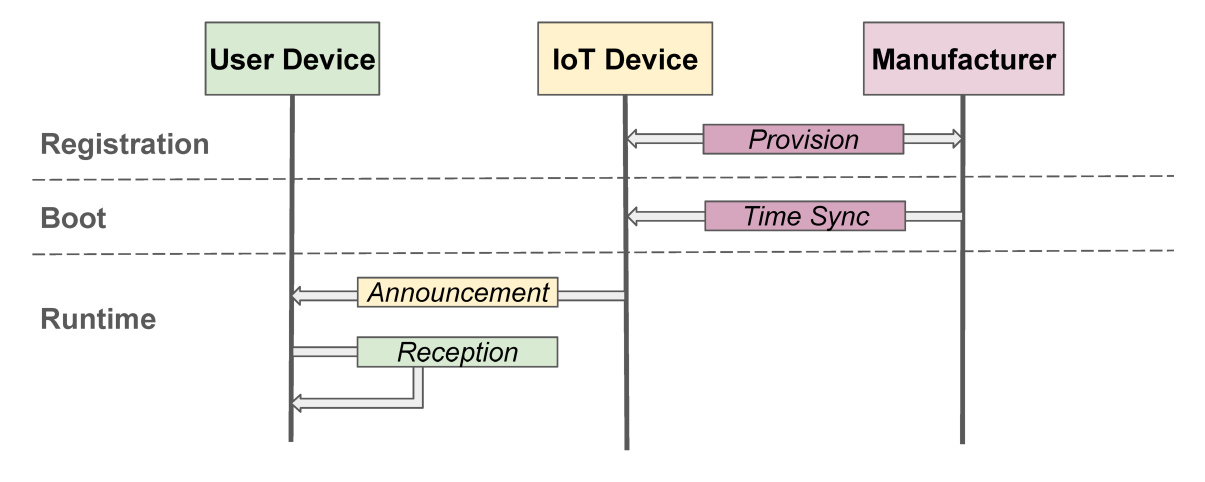}}
	\subfigure[\paisaplus \pull Model Overview \label{fig:pull}]
	{\includegraphics[width=0.49\textwidth]{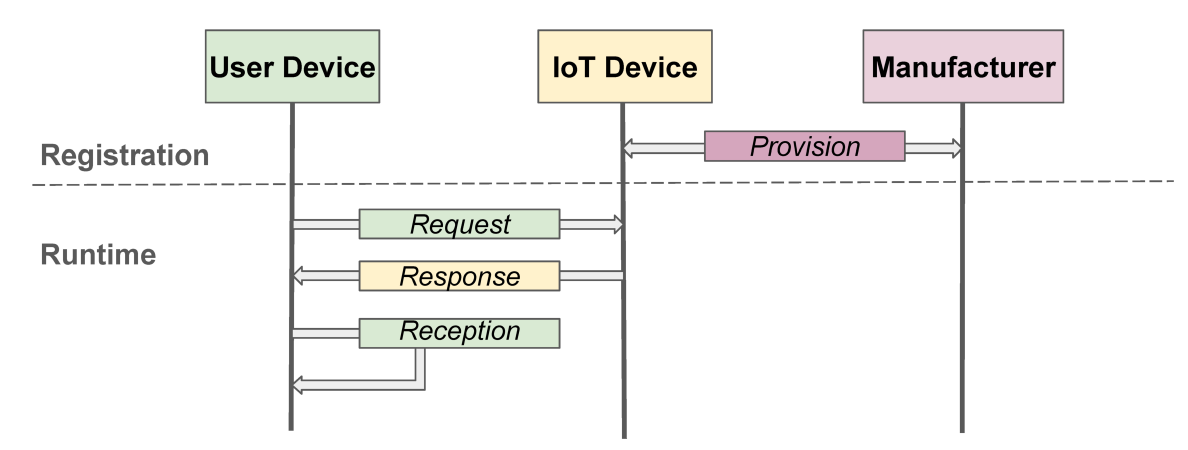}}
      \vspace*{-0.2cm}
	\caption{\push and \pull Models Overview}
     \vspace*{-0.1cm}
	\label{fig:models-overview} 
\end{figure*}

Similar to prior work, all messages exchanged between \iotdev and \usrdev are broadcasted without
any prior security associations or channel/session establishment.
As mentioned earlier, in the \push model of \paisa \cite{paisa} and DIAL \cite{dial}, \iotdev periodically 
announces itself, and \usrdev (if present) receives and processes such information from
nearby devices. 
Conversely, in \pub, the user initiates the process via an app on \usrdev.
\usrdev broadcasts a discovery request and waits.
Upon receiving a request, \iotdev generates a response.
Figure \ref{fig:models-overview} depicts an overview of \push and \pull models.
Unlike \paisa, which needs a time synchronization phase, \pub has only two phases: \registration and \runtime.

\noindent {\bf $\boldsymbol{\registration}$ Phase} takes place during the manufacturing of \iotdev.
\mfr installs software and provisions secrets for the underlying public key 
algorithm (i.e., \iotdev key-pair) as well as some metadata.

\noindent {\bf $\boldsymbol{\runtime}$ Phase} has three steps: \request, \response, and \receptionusr.
Once completing the boot sequence, \iotdev keeps listening for requests while performing normal operations.
In \request step, \usrdev broadcasts a request (\requestmessage) to solicit device information from nearby \iotdev-s 
and waits.
Upon receiving \requestmessage, \iotdev generates and returns a response (\responsemessage) in \response step.
Next, \usrdev processes and verifies \responsemessage in \receptionusr step, and displays 
\iotdev information to the user. A detailed description of this protocol is in Section \ref{sec:pull_detail}.

\subsection{Adversary Model}\label{subsec:pulladversarymodel}
An adversary (\sadv) can access all memory regions (e.g., flash and RAM), except for the 
TCB (defined in Section \ref{subsec:reg}) and its data within the TEE.
Outside the TCB, any \iotdev components and peripherals, including timers, 
network interfaces, sensors, and actuators, are subject to compromise.
Communication between \iotdev-s and \usrdev-s is subject to eavesdropping and manipulation by 
\sadv following the Dolev-Yao model \cite{DolevYao}. Additionally, \registration phase is 
considered to be secure. \mfr is trusted to accurately provision IoT devices and safeguard their secrets.
Similarly, \pub app on \usrdev is considered trustworthy. Though it maintains no secrets, it is 
assumed to correctly generate formatted requests as well as accurately process and 
display information to the user.

\noindent {\bf Denial-of-Service (DoS):}
\sadv can exploit vulnerabilities in non-TCB software to introduce malware, and then 
attempt to exhaust \iotdev's computing resources, e.g., cores, memory, and peripherals. 
Through such malware, it can also occupy (squat on) 
non-TCB-controlled peripherals from inside the device.
Also, it can swamp peripherals from the outside, e.g., via excessive network traffic or fake 
request flooding, which is possible since requests are not authenticated.
Sections \ref{subsec:pull_challenge} and \ref{subsec:discussion-dos} describe how
\pub fully mitigates malware-based, and partially defends against network-based, attacks.

\noindent {\bf Replay Attacks:}
\sadv can reply with stale, yet valid, responses to \usrdev.
Replay attacks on \iotdev are not a serious concern due to lack of authentication for \requestmessage.
Replays of stale \requestmessage-s are a special case of request flooding mentioned above.

\noindent {\bf Wormhole Attacks:}
\pub does not defend against wormhole attacks \cite{wormhole4}. In such attacks,
\sadv tunnels both \requestmessage and its corresponding \responsemessage to/from a remote\footnote{Here, 
``remote'' means: beyond one-hop range of a \usrdev.} \iotdev, thus fooling a \usrdev into 
believing that a far-away \iotdev\ is nearby. Well-known prior techniques 
\cite{wormhole1,wormhole2,wormhole3,surveydistancebounding} can address wormhole attacks.
However, mounting a wormhole attack in \pub\ is strictly harder than in \paisa, since \sadv
must tunnel both \requestmessage and \responsemessage, while only the latter suffices in \paisa.

\noindent {\bf Runtime Attacks:}
Similar to \paisa, \pub detects software modifications via periodic \ra.
However, runtime attacks (e.g., control-flow and non-control-data) are out of scope.
Prior research \cite{acfa,litehax,cflat,tinycfa,oat,embedded-CFI} has proposed various 
mitigation techniques, such as control-flow attestation (CFA) and control-flow integrity (CFI).
Unfortunately, these techniques are resource-intensive and thus usually impractical for
lower-end IoT devices.

\noindent {\bf Physical Attacks:}
\pub protects against non-invasive physical attacks, e.g., reprogramming \iotdev 
using a legal interface, such as JTAG.
However, \pub does not protect against physically invasive attacks, such as 
hardware faults, ROM manipulation, or secret extraction via side-channels \cite{zankl2021side}. 
For defenses against these attacks, we refer to \cite{ravi2004tamper}.
 
\noindent {\bf Non-compliant (Hidden) Devices:}
We do not consider attacks whereby \sadv gains physical access to an environment 
and surreptitiously deploys non-compliant malicious devices. For this purpose,
we refer to some recent research on detecting and localizing hidden devices via 
specialized hardware \cite{mmwave,NLJD,bugdetector,sengupta2020mm} and network traffic 
analysis \cite{lumos,snoopdog}.

\noindent {\bf Co-existence with Other Secure Applications:}
In single-enclave TEEs (e.g., ARM TrustZone), all secure resources
are shared by all secure applications. Consequently, a compromised secure application 
can access \pub secrets or mount a DoS attack by squatting on the network interface, which is part of \pub TCB.
As in \paisa, we assume that all secure applications are free of vulnerabilities.
Although this issue can be addressed by TEEs that support multiple enclaves (e.g., 
Intel SGX \cite{sgx}, ARM CCA \cite{cca}), such TEEs are generally unavailable on 
IoT devices targeted by \pub.

\subsection{Requirements}
Recall that \pub aims to facilitate {\it efficient} and {\it guaranteed} timely responses 
on \iotdev-s upon requests from nearby \usrdev-s.
In terms of performance and functionality, \pub must satisfy the following properties:
\begin{compactitem}
    \item {\it Low latency:}
    \requestmessage reception and \responsemessage generation time on \iotdev must be minimal. 
    \pub implementation should have a minimal impact on \iotdev's normal operation.
    \item {\it Low bandwidth:}
    \pub messages (\requestmessage and \responsemessage) must consume minimal bandwidth.
    \item {\it Scalability:}
    \iotdev should be able to handle multiple \requestmessage-s and respond in time.
    \item {\it Casualness:} 
    As in \paisa, no prior security association between \usrdev-s and \iotdev-s should be assumed, and
    secure sessions/handshakes between them must not be required.
\end{compactitem}
To mitigate attacks defined in Section \ref{subsec:pulladversarymodel}, \pub must provide:
\begin{compactitem}
    \item {\it Unforgeability:} \usrdev must verify that \responsemessage comes from a
    genuine \pub-enabled \iotdev, i.e., \sadv cannot forge \responsemessage.
    \item {\it Timeliness:} \usrdev must receive \responsemessage containing \iotdev 
    information in a timely fashion. 
    \item {\it Freshness:} \responsemessage must be fresh and reflect
    the recent software state of \iotdev.
\end{compactitem}

\subsection{DoS Attacks} \label{subsec:pull_challenge}
We now consider DoS attacks on \usrdev and \iotdev.

A network-based \sadv can mount DoS attacks by flooding
\usrdev with fake \responsemessage-s. This forces \usrdev 
to verify numerous (invalid) signatures, clearly wasting resources. 
\pub offers no defense against such attacks, due to the casual nature of 
the \iotdev-\usrdev relationship.
Also, this may not be a critical issue on higher-end \usrdev.

As discussed in Section \ref{subsec:pulladversarymodel}, there are two types of 
DoS attacks on \iotdev: (1) an internal malware-based \sadv who keeps the 
CPU and/or network peripherals busy, and (2) a network-based \sadv who floods \iotdev 
with frivolous \requestmessage-s and thereby depletes computing resources with expensive 
cryptographic operations needed for \responsemessage generation.
Similar to \paisa, (1) is mitigated by configuring the TEE such that a network 
peripheral is placed under the exclusive control of the TCB. 
Consequently, \requestmessage reception and \responsemessage generation tasks have the highest priority.

Dealing with (2) is more challenging. Since \requestmessage-s are not authenticated, 
a flood of fake \requestmessage-s can simply exhaust 
\iotdev resources. The same can occur in a non-hostile scenario when a flurry 
of valid requests from multiple benign users (e.g., in a suddenly crowded space) overwhelms \iotdev.

To mitigate \requestmessage flooding (whether hostile or not), \pub uses a 
lazy-response technique whereby \iotdev collects \requestmessage-s for
a certain fixed time and responds to all of them with a single \responsemessage. 
Specifically, \iotdev maintains a pool of nonces (up to a certain maximum number)
from \requestmessage-s. At the end of the period or when the nonce pool reaches its capacity, 
\iotdev signs a collective 
\responsemessage containing all pooled nonces. Each \usrdev with an outstanding 
\requestmessage checks whether the received \responsemessage contains its nonce. 
This approach allows \iotdev to pace its compute-intensive cryptographic operations. 
Details are in Section \ref{sec:pull_detail}.

We acknowledge that pooling nonces from \requestmessage-s does not fully
address \requestmessage flooding: \sadv can simply flood \iotdev at a high
enough speed, causing \iotdev to generate signed \responsemessage-s at an
unsustainable rate. One remedial measure is to adopt random deletion -- a relatively 
effective countermeasure to the well-known TCP SYN-flooding attack \cite{schuba1997analysis}.
Furthermore, nonce pooling delays \responsemessage generation, making \usrdev-s wait longer. 
To address this issue, an alternative approach (albeit with a slightly different setting) is discussed in Section \ref{subsec:discussion-dos}.

\section{\pub Design}\label{sec:pull_detail}
Recall that \pub has two phases: \registration and \runtime, 
as shown in Figure \ref{fig:pull}.

\subsection{Registration Phase}\label{subsec:reg}
In this phase, \mfr provisions each \iotdev with device software (\iotdevsoft), 
\pub trusted software (\pubtcb), cryptographic keys, and other metadata.
Also, the timer and network peripherals are configured as secure via the TEE.
Finally, \mfr configures the attestation interval (\attestinterval) and the 
lazy-response delay (\geninterval) according to each \iotdev use case.
Their use is described in Section \ref{subsec:runtime}.

\noindent {\bf Software:}
Normal functionality of \iotdev is managed by \iotdevsoft, which resides in
and executes from the non-secure memory region. Meanwhile,
\pubtcb\ is installed in the secure region.

\noindent {\bf Cryptographic Keys:}
There is a public/private key-pair for \iotdev (\iotdevpk, \iotdevsk). The latter is 
used to sign \responsemessage. This key pair is generated inside the TEE.
\iotdevpk\ is shared with \mfr, while \iotdevsk\ never leaves the TEE.
\mfrpk\ is assumed to be trusted, e.g., via a public key infrastructure (PKI) dealing with all \mfrpk-s.

\noindent {\bf Device Manifest ($\boldsymbol{\devmanifest}$)}
is the information about \iotdev, generated and maintained by \mfr.
It must include \mfr certificate (\certmfr) and \iotdev certificate (\certiot) 
signed by \certmfr. The exact contents of \devmanifest are up to the individual \mfr. 
However, some fields are mandatory: \iotdevpk, certificates, type/model of \iotdev, 
sensors/actuators it hosts, etc. Note that \devmanifest 
is not placed into \iotdev; it is hosted by \mfr at a URL, \devmanifesturl.

Once \usrdev obtains \devmanifest, it uses \certmfr to authenticate 
it, and extracts \iotdevpk \ from \certiot. \devmanifest might contain 
other device information, e.g., purposes of its sensors/actuators, the specification 
link, \iotdevsoft\ version, and coarse-degree deployment geographical location, 
e.g., country, state, or city.\footnote{This coarse location information can aid 
in (partially) mitigating wormhole attacks.}

\noindent {\bf URL ($\boldsymbol{\devmanifesturl}$):}
\devmanifest size can be large since it depends on \mfr and type/model of \iotdev. Thus,
we cannot expect it to fit into one \responsemessage. Therefore, \devmanifest is indirectly
accessible at a shortened URL, called \devmanifesturl, contained in each \responsemessage. 
\devmanifesturl is a shortened version of \devmanifesturlfull, created using a
URL shortening service, such as \verb+Bitly+\cite{bitly} or \verb+TinyURL+\cite{tinyurl}, 
for making it short and of constant size.

\noindent {\bf Metadata:}
The following parameters are stored in the secure memory region of \iotdev:
(1) \devmanifesturl, (2)  \iotdevsofthash -- hash of \iotdevsoft, 
(3) \attestinterval -- inter-attestation time parameter, (4) \geninterval -- \responsemessage 
generation interval, and (5) \poolsize -- maximum size of the nonce pool. 

\noindent {\bf TCB:}
Cryptographic keys, \pubtcb, and all aforementioned metadata (stored in the
secure memory region) are considered to be within the TCB, along with the timer and
network interface, which are configured as secure peripherals.

\subsection{Runtime Phase}\label{subsec:runtime}
\runtime phase composed of three steps: \request, \response, and \receptionusr.

\begin{figure}[t]
  \centering  
  \captionsetup{justification=centering}  
  \includegraphics[width=0.95\columnwidth]{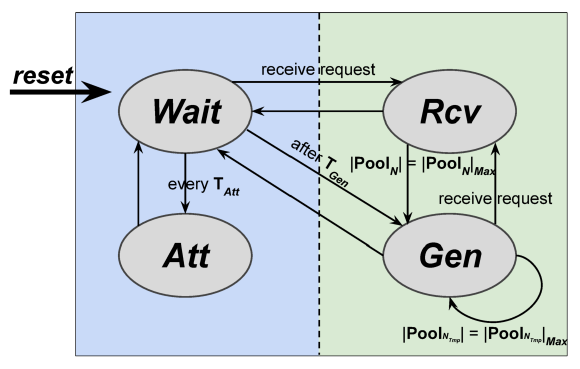}
  \vspace*{-.3cm}  
  \caption{\pub State Machine on \iotdev}
  \label{fig:pull_state_machine}
  \vspace*{-.5cm}  
\end{figure}

\subsubsection{\pub Trusted Software (\pubtcb) on \iotdev}
As shown in Figure \ref{fig:pull_state_machine}, \iotdev has four states in \response step: 
\wait, \attest, \rcv, and \gen. \attest is a periodic state independent
of others: its periodicity is governed by \attestinterval.

\noindent{\bf (a) $\boldsymbol{\wait}$:}
After its boot sequence completes, \iotdev runs \iotdevsoft\ while listening 
for \requestmessage-s on the network interface. Depending on the condition below, 
it transitions to other states:
\begin{compactitem}
    \item {\it Every \attestinterval}: \iotdev periodically attests \iotdevsoft\ via a 
    secure timer set off
    every \attestinterval. When the timer 
    expires, \iotdev proceeds to $\boldsymbol{\attest}$.
    \item {\it \requestmessage received}: Once \iotdev identifies an incoming packet as a 
    \requestmessage (via \pub protocol identifier -- \pubid, in a header field), it transitions to 
    $\boldsymbol{\rcv}$.
    \item {\it After \geninterval}: It proceeds to $\boldsymbol{\gen}$.
\end{compactitem}

\noindent{\bf (b) $\boldsymbol{\attest}$:}
\iotdev computes a hash over \iotdevsoft\ and compares it to \iotdevsofthash.
If they do not match, the 1-bit \attestresult\ flag is set to \verb+Fail+ and 
to \verb+Success+ otherwise. Next, \iotdev securely records its current timer 
value (\tsvalue{L_{\attest}}) and returns to $\boldsymbol{\wait}$.

An intuitive alternative to timer-based attestation is to perform attestation 
upon each \requestmessage. The attestation itself consumes far less time/energy than
signing \responsemessage\.
However, for a low-end \iotdev\ that performs safety-critical tasks, 
decoupling attestation from discovery can be more appealing. 

\noindent{\bf (c) $\boldsymbol{\rcv}$:} 
As discussed in Section \ref{subsec:pull_challenge}, \pub uses a flexible lazy-response strategy 
to amortize signature costs and mitigate potential DoS attacks via \requestmessage\ flooding. 
In this state, \iotdev maintains \nlist, composed of nonces extracted from \requestmessage-s.

\begin{compactitem}   
    \item \iotdev initially transitions to this state when it receives the first \requestmessage.
    It extracts the nonce from \requestmessage, \noncevalue{\usr}{}, and places it 
    into a (initially empty) nonce pool \nlist.
    It then sets a secure timer to \geninterval and returns to $\boldsymbol{\wait}$. Note that the timer for \geninterval is only set when the first \requestmessage comes in.
    \item If \nlist is not empty, \iotdev extracts the nonce and adds it to \nlist.
    Once \nlist reaches \poolsize, \iotdev\ transitions to $\boldsymbol{\gen}$.
    Otherwise, it returns to $\boldsymbol{\wait}$.
    \item If it has been transitioned from \gen (i.e., \requestmessage arrives while generating \responsemessage), \iotdev extracts the nonce and adds it to a temporary list -- \ntmplist, and returns to \gen.
\end{compactitem}
\noindent Although \poolsize is a configurable parameter, it is physically upper-bounded by
the amount of space available in \responsemessage, which depends on the network interface and 
the underlying wireless medium. This is further discussed in Section \ref{subsec:imple-pull}.

Note that if \poolsize is set to $1$, a separate
\responsemessage\ is generated for each \requestmessage.
Also, \poolsize should not be set over the physical upper bound because it leads to all \requestmessage-s ignored until the current \responsemessage is completely handled.
This can result in a false sense of privacy, frustrating users to erroneously think there are no \iotdev-s around.

\noindent{\bf (d) $\boldsymbol{\gen}$:}
\iotdev computes attestation time (\tsvalue{\attest}) as:
\[\tsvalue{\attest} = \tsvalue{\iot} - \tsvalue{L_{\attest}}\]
where \tsvalue{\iot} is the current timer value.
Then, it generates $\attestreport:=[\attestresult,\tsvalue{\attest}]$, signs the message 
$\sigvalue{\resp} := \sigfunc(\noncevalue{dev}{} || \nlist \\ || \devmanifesturl || \attestreport$), 
and creates $\responsemessage:=[\noncevalue{dev}{},\nlist,\devmanifesturl,\\ \attestreport,\sigvalue{\resp}]$.
\sigfunc \ is a signing operation and 
\noncevalue{dev}{}, generated by \iotdev, is added to randomize \responsemessage.
Next, \iotdev assigns \ntmplist to \nlist, resets \ntmplist to be empty, and broadcasts \responsemessage.
Note that if an influx of \requestmessage-s (i.e., \# of \requestmessage-s $> 2*$\poolsize) is received within \geninterval, \ntmplistsize exceeds \poolsize.
Then, the first block (of length \poolsize) of \noncevalue{usr}{}-s from \ntmplist is moved to \nlist after the current \responsemessage is sent out.
The remaining \noncevalue{usr}{}-s in \ntmplist are in turn handled similarly when \gen is triggered.
Finally, it goes to the next state with the below conditions:
\begin{compactitem}   
    \item If \ntmplistsize < \poolsize, it transitions to $\boldsymbol{\wait}$.
    \item Otherwise, it re-enters $\boldsymbol{\gen}$ to process a new \responsemessage with next \noncevalue{usr}{}-s.
\end{compactitem}
If any \requestmessage-s are received in this state, it goes to \rcv to handle them.

\subsubsection{\pub App on \usrdev} 
There are two steps: \request \& \receptionusr.

\noindent{\bf $\boldsymbol{\request}$:}
\usrdev initiates device discovery by broadcasting a \requestmessage that contains a unique
\noncevalue{\usr}{} and \pubid. It then waits for \responsemessage-s for a time period set as 
part of the \pub app configuration. 

\noindent{\bf $\boldsymbol{\receptionusr}$:}
Upon reception of \responsemessage, \usrdev parses it and checks for the presence of
\noncevalue{\usr}{}. If not found, \responsemessage\ is discarded.
Next, \usrdev fetches \devmanifest from \devmanifesturl.
It then retrieves \mfrpk\ from \certmfr\ and verifies the signature of 
\devmanifest using \mfrpk. Successful verification implies that both \devmanifest\ and \iotdevpk\ are 
trustworthy. Finally, \usrdev verifies \sigvalue{\resp} with \iotdevpk\ and 
displays the details to the user.

\section{Implementation}\label{sec:implementation}
This section describes \pub implementation details. All source code is 
available at \cite{paisaplus-code}. 

\begin{figure}[t]
  \centering
  \captionsetup{justification=centering}
  \includegraphics[width=0.95\columnwidth]{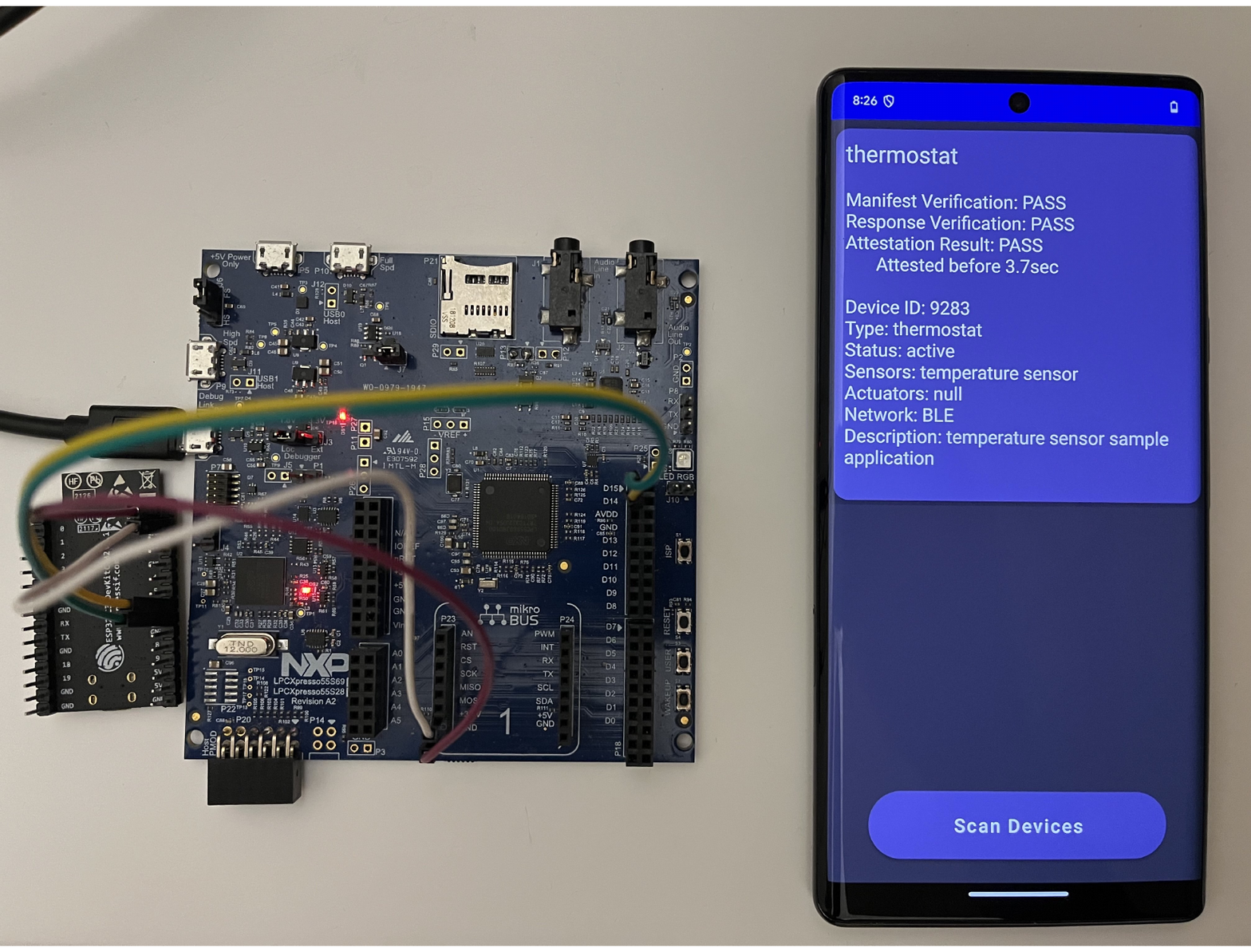}
  \vspace{-0.1cm}
  \caption{Implementation Setup}
  \label{fig:imp-setup}
  \vspace{-0.4cm}
\end{figure}

\subsection{Implementation Setup}\label{subsec:impl_setup}
\iotdev is implemented on an NXP LPC55S69-EVK \cite{nxpboard} board.
It features an ARM Cortex-M33 processor with ARM TrustZone-M (TZ-M), running at 
$150$ MHz with $640$ KB flash and $320$ KB SRAM.
Due to lack of a built-in network module, we use an ESP32-C3-DevKitC-02 board \cite{espboard}, 
connected to the NXP board via UART. As \usrdev, we use a 
Google Pixel 6 Pro Android smartphone \cite{pixel6-pro} with eight heterogeneous cores running at 
$\leq~2.8$ GHz. Figure \ref{fig:imp-setup} shows the overall experimental setup.

\noindent {\bf Secure Peripherals:}
We configure UART4 and CTIMER2 as secure, granting \pubtcb\ exclusive control over the network and timer peripheral.
UART4 and CTIMER2 are configured with the highest priority values (0 and 1, respectively)
to ensure timely and guaranteed reception of \requestmessage-s and generation of \responsemessage-s.
If \iotdevsoft\ needs to access the network peripheral to communicate with an external entity, 
it can use UART4 for its normal operation via \pubtcb.
\attestinterval is set to $300$s and \geninterval\ -- to $1$s on CTIMER2.
Also, CASPER and HASH-AES Crypto Engine peripherals facilitate 
cryptographic operations, i.e., signing \responsemessage-s and 
hashing (non-secure) memory during attestation.

\begin{figure}[t]
  \centering
  \captionsetup{justification=centering}
  \includegraphics[width=\columnwidth]{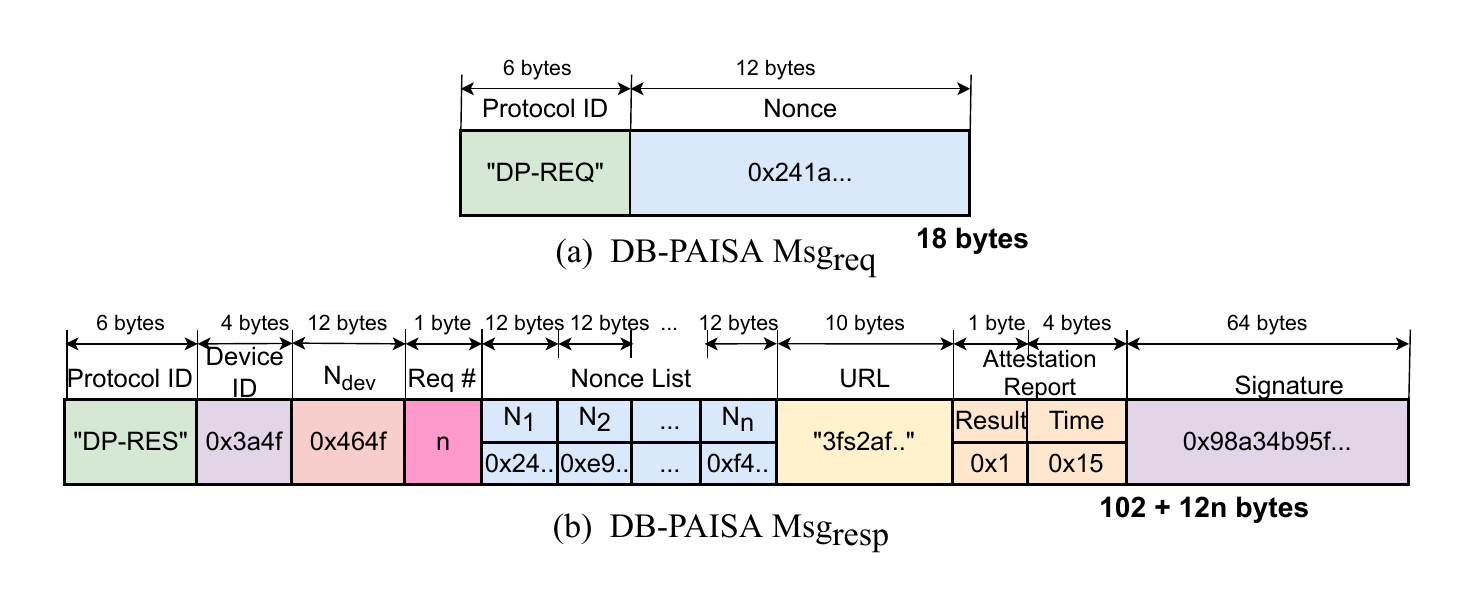}
  \vspace{-.8cm}
  \caption{\requestmessage and \responsemessage Formats \\ (n: \# of requests within \geninterval, and n $\geq 1$)}
  \label{fig:pub-msg}
  \vspace{-0.3cm}
\end{figure}

\subsection{Network}\label{subsec:impl-network}
\noindent {\bf Device Discovery:}
We focus on popular wireless media types: WiFi and Bluetooth.
Each has a network discovery procedure. To initiate discovery, one typically broadcasts a 
short (specially formatted) network discovery message. 
Most fields in this message are reserved and of fixed size.
Thus, using network discovery messages as \requestmessage-s and \responsemessage-s 
is not trivial.

\noindent {\bf $\boldsymbol{\requestmessage}$ \& $\boldsymbol{\responsemessage}$ Length:}
As shown in Figure \ref{fig:pub-msg}, \requestmessage size is constant -- $18$ bytes. 
The minimum size of a \responsemessage is $114$ bytes for a single nonce. Since one
\responsemessage can reply to multiple \responsemessage-s (each with its own nonce), 
we need to encode the total number of nonces in front of the nonce list. 
We use one extra leading byte for this purpose.
In general, a network discovery packet must allow for at least $114$ 
bytes of \pub-specific data.

\noindent {\bf Bluetooth {\em vs} WiFi:}
Both WiFi and Bluetooth are common on many types of IoT devices as well as smartphones,
smartwatches and tablets. 
To support lightweight connection-less communication between \iotdev-s and \usrdev-s, 
our \pub prototype uses Bluetooth extended advertisements.
The rationale for this choice is three-fold:
(1) Bluetooth 5 supports broadcast messages of $\leq~1,650$ bytes,
while IEEE 802.11 WiFi can only carry $\leq~255$ bytes in a vendor-specific field of a beacon frame,
(2) it is more energy-efficient than WiFi \cite{blevswifi}, and
(3) its indoor range of $\approx 40$m is more appropriate for \pub, 
since \iotdev-s discovered via WiFi may be irrelevant to a \usrdev due to being too far.

\subsection{Normal Operation on \iotdev (\iotdevsoft)}
We implemented a temperature sensor application as \iotdev's normal functionality -- \iotdevsoft.
The same application was used to motivate and evaluate \paisa.
It obtains temperature sensor readings on \iotdev using LPADC (Low-Power Analog-to-Digital Converter) 
driver every $5$s, and sends the data to a remote server.
Due to UART4 being set as secure, \iotdevsoft\ cannot use it directly; it first goes through 
\pubtcb, running in Secure world, to send packets.
This is implemented using a Non-Secure Callable (NSC) function, which is the only valid entry point 
to transition from Normal world to Secure world in TZ-M, except for secure interrupts.
\iotdevsoft\ is implemented as a simple task on freeRTOS \cite{freeRTOS}, reading the 
temperature sensor with $5$s delay in a loop.

\subsection{\pub Trusted Software} \label{subsec:imple-pull}
\pub includes three trusted applications: (1) secure application (\pubtcb) running in 
Secure world on \iotdev, (2) network stack connected to \iotdev via secure 
UART4, and (3) Android app on \usrdev. 

\noindent {\bf $\boldsymbol{\pubtcb}$ on $\boldsymbol{\iotdev}$:}
\pub uses a secure timer in three cases:  (1) triggering an interrupt 
at \attestinterval ($300$s) to perform attestation, (2) triggering an interrupt at 
\geninterval ($1$s) for the lazy-response mechanism, and (3) computing
estimated attestation time. The first two require the timer interrupt to be triggered 
at different intervals. Fortunately, most commercial timers support multiple conditions 
triggering the interrupt. Hence, \pubtcb\ requires only one exclusive secure timer.

\pubtcb \ has four software components: UART interrupt service routine (ISR), 
timer ISR, attestation, and \responsemessage generation.
Once it boots, \iotdev continuously listens for \requestmessage-s.
UART ISR is triggered whenever \iotdev receives a packet from the network module.
It identifies \requestmessage-s by \pubid\ (\verb+"DP-REQ"+) 
and prioritizes handling them. If \nlist is empty, it places \noncevalue{\usr}{} 
into \nlist\ and sets the secure timer to expire after \geninterval.
Otherwise, it adds (appends) \noncevalue{\usr}{} to \nlist.

As mentioned earlier, two conditions can trigger timer ISR: \attestinterval and \geninterval.
This timer ISR checks which timer expired. If it is \attestinterval, ISR 
computes \verb+SHA256+ over non-secure memory (including \iotdevsoft) and generates \attestresult.
Otherwise, if \geninterval expired, timer ISR initiates \responsemessage generation process, which
(1) computes \tsvalue{\attest}, (2) signs 
\responsemessage contents (\noncevalue{dev}{}, \nlist, \devmanifesturl, and \attestreport) using 
\verb+ECDSA+ (\verb+Pri-+ \verb+me256v1+ curve), (3) composes \responsemessage, and (4) 
hands it over to the network module via \verb+UART_WriteBlocking()+.
Note that, in order to prioritize \responsemessage generation, the timer interrupt from \attestinterval is 
inactivated when \geninterval is reached.
Also, when \nlist is empty, the timer interrupt from \geninterval is {\bf not} in use, i.e., not set.

When \nlist reaches its capacity (\poolsize), \responsemessage generation is triggered.
\responsemessage generation can also be triggered by \geninterval expiring.
UART ISR has the highest priority to receive \requestmessage-s 
and process \noncevalue{\usr}{}-s, even during \responsemessage generation.
If UART ISR receives a new \requestmessage while \responsemessage is being generated,
\noncevalue{\usr}{} from this new \requestmessage is stored in a separate temporary list, \ntmplist.
It is fed into \nlist after handling the current \responsemessage.

The current minimum nonce length recommended by NIST 
for lightweight cryptography~\cite{nist-nonce} is $12$ bytes.
In Bluetooth 5, a single \responsemessage can contain $\leq~129$ 
nonces, meaning that it can collectively respond to as many \requestmessage-s. This upper bound is 
based on the maximum capacity of Bluetooth extended advertisements ($1,650$ bytes)
and other \responsemessage fields, e.g., $64$-byte signature, and $5$-byte \attestreport.

\noindent {\bf Network Module:} ESP32-C3-DevKitC-02 board features Bluetooth 
5, which supports Bluetooth extended advertisements. To send/receive such broadcast 
messages (\requestmessage, \responsemessage), we use the \verb+NimBLE+ library. 
Once the boot sequence completes, the network module initializes and 
begins scanning for \requestmessage-s using \verb+ble_gap_disc()+.
Upon receiving \requestmessage, the module forwards it to the main board (NXP board),
using \verb+uart_write_bytes()+. 
When the NXP board replies with \responsemessage,
the network module broadcasts out \responsemessage to \usrdev, using \verb+ble_gap_ext_+ \verb+adv_start()+.
The network module (re)transmits \responsemessage every $30$ms for $300$ms 
to ensure reliable delivery. These timing parameters are configurable.

\noindent {\bf $\boldsymbol{\pub}$ App on $\boldsymbol{\usrdev}$:}
\pub app is implemented as an Android app on \usrdev, using Android Studio Electric Eel with API level 33. 
For a Bluetooth scan and broadcast, the app requires a few 
permissions: \verb+BLUETOOTH_SCAN+, \verb+BLUETOOTH_ADVERTISE+, 
and \verb+ACCESS_FINE_LOCATION+, which are reflected in `AndroidManifest.xml' file.

The app uses \verb+android.bluetooth.le+ libraries for Bluetooth extended advertisements. 
When the user clicks the ``Scan Devices'' button, the app broadcasts \requestmessage using 
\verb+startAdvertisingSet()+ and waits with a scan for \responsemessage-s using \verb+startScan()+.
\responsemessage is identified by containing \pubid\ (\verb+"DP-RES"+). 
The app parses \responsemessage and fetches \devmanifest via \devmanifesturl using 
\verb+getInputStream()+ in the \verb+java.net+ library. 
Then, the app verifies signatures in \devmanifest and \responsemessage with 
\mfrpk \ and \iotdevpk, using \verb+verify()+ from the \verb+java.+ \verb+security+ 
library. Finally, it displays device information details (from \devmanifest) and \attestreport 
(from \responsemessage) on \usrdev, as shown in Figure \ref{fig:imp-setup}.

\section{Evaluation}\label{sec:evaluation}
\pub security is addressed in Section \ref{subsec:security-analysis}.
For the sake of a fair comparison, since \paisa was originally implemented using WiFi, 
we adapt it to Bluetooth. We measure runtime overhead on \iotdev and \usrdev 
across 50 iterations. \response on \iotdev takes $233$ms, consistent with the time 
\paisa announcement takes. We use this value to discuss performance overhead and energy consumption 
in Sections \ref{subsec:eval-perf} and \ref{subsec:eval-energy}, respectively.
We also empirically measure the \pub-imposed performance penalty of \iotdevsoft\ 
with varying parameters in Section \ref{subsec:eval-perf}.
Runtime overhead on \usrdev and overall network traffic overhead are evaluated in 
Sections \ref{subsec:eval-run-usr} and \ref{subsec:eval-network-overhead}.

\subsection{Security Analysis}\label{subsec:security-analysis}
We now consider security guarantees of \pub\ against \sadv 
defined in Section \ref{subsec:pulladversarymodel}.

\noindent{\bf $\boldsymbol{\iotdev}$ Compromise:} TZ-M guarantees that the 
TCB cannot be manipulated by \sadv.
Also, due to the highest priority of UART and timer interrupts, \iotdev is guaranteed 
to perform \pub tasks (i.e., receiving, processing, and replying to \requestmessage-s) even 
with malware presence in TZ-M Normal world.
Furthermore, \attestreport contained in \responsemessage reflects the recent
state of \iotdevsoft\ which allows \usrdev-s to detect compromised \iotdev-s.

\noindent{\bf Forged $\boldsymbol{\responsemessage}$:}
\sadv cannot generate a valid \responsemessage unless \iotdevsk\ 
is leaked or the public key algorithm used for digital signature computation is broken.

\noindent{\bf DoS Attacks on $\boldsymbol{\iotdev}$}
can be launched by (1) malware on \iotdev or (2) other malicious devices over the network.
Since timer ISR and UART ISR are configured as secure, the former can be addressed with 
\pubtcb's exclusive control over the network interface. The latter requires the physical presence of
\sadv within Bluetooth range of \iotdev, making this type of attack harder.
This is partially addressed with the lazy-response approach
as discussed in Sections \ref{subsec:pull_challenge} and \ref{subsec:runtime}.
Other mitigation techniques are outlined in Section \ref{subsec:discussion-dos}.

\noindent{\bf Replay Attacks on $\boldsymbol{\usrdev}$:}
\usrdev can readily detect replay attacks by checking whether \noncevalue{\usr}{} from an outstanding 
\requestmessage is included in corresponding \responsemessage. Such \responsemessage-s can be simply discarded.

\noindent{\bf Physical Attacks on $\boldsymbol{\iotdev}$:}
TZ-M offers secure storage to store secrets and secure boot to thwart non-invasive physical attacks.

\subsection{\iotdev Runtime Overhead} \label{subsec:eval-perf}
To measure runtime overhead on \iotdev, we use \busytime\ to denote CPU usage for \pub trusted 
software (\pubtcb), computed as: $\busytime := \frac{\tsvalue{\pub}}{\tsvalue{Normal} + \tsvalue{\pub}}$, 
where \tsvalue{Normal} is time used for normal device operation (\iotdevsoft) and \tsvalue{\pub} 
is time used by \pubtcb. For the sake of simplicity, despite the presence of
periodic attestation every \attestinterval, only \responsemessage generation overhead is considered 
in the experiment because attestation time is almost negligible ($\approx1$ms) compared to 
its interval (\attestinterval), which can be quite long, e.g., $1$ hour.
Also, the signing operation ($231$ms out of $233$ms) dominates \responsemessage generation.
Therefore, \responsemessage generation takes (almost) constant time even with a barrage of 
\requestmessage-s within one \geninterval in our lazy-response design.

In \paisa \push model, \busytime can be represented as: $\busytime := \frac{\tsvalue{Ann}}{\anninterval}$,
where \tsvalue{Ann} is time to generate an announcement and \anninterval is inter-announcement
interval. Since \tsvalue{Ann} is constant ($\approx 232$ms), \busytime is a function of  
configurable \anninterval.

\begin{figure}[t]
  \vspace*{-.8cm}  
  \centering  
  \captionsetup{justification=centering}  
  \includegraphics[width=.95\columnwidth]{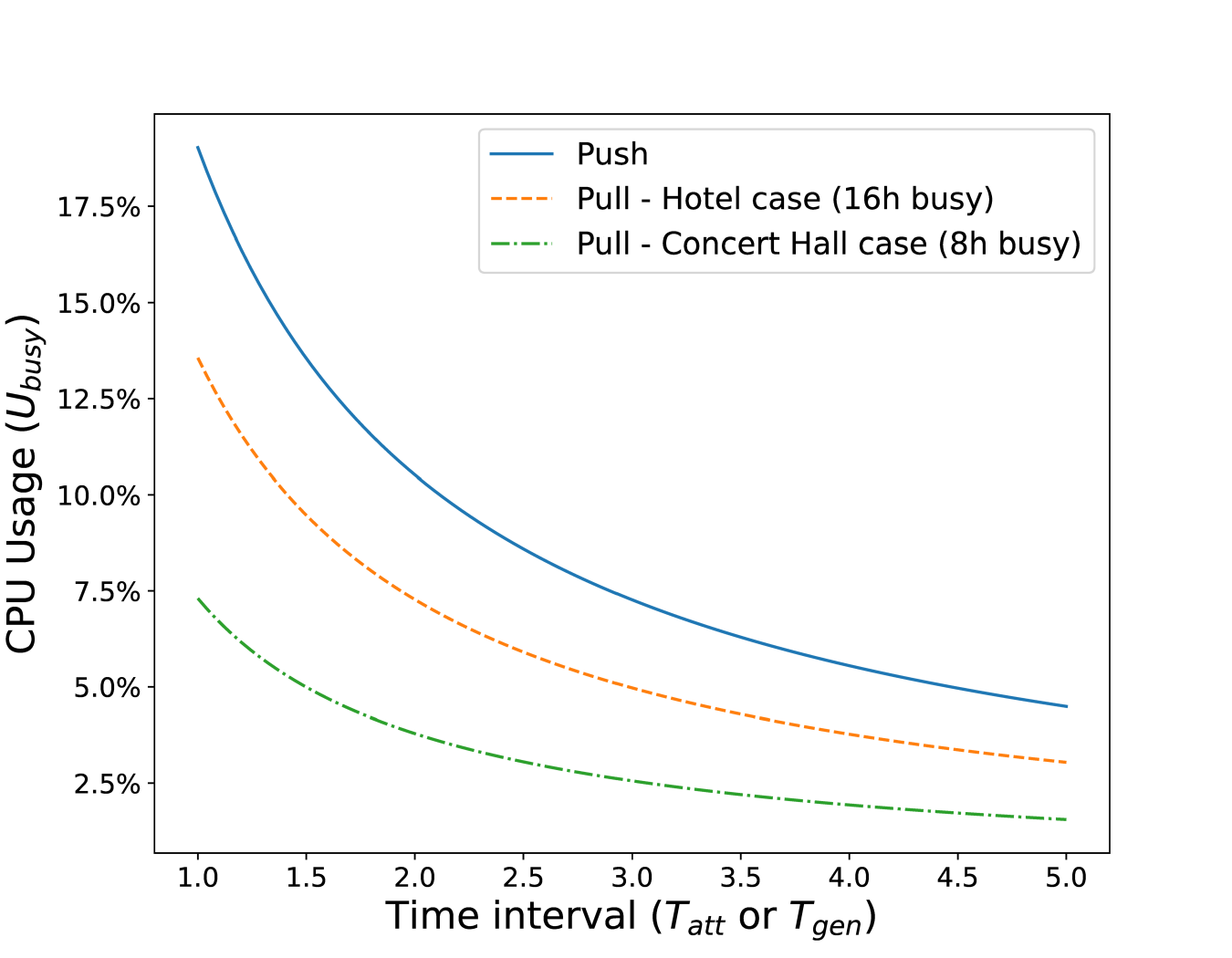}
  \vspace*{-.3cm}  
  \caption{CPU Usage (\busytime) in (\dbb)\paisa \push and \pull models \\ (Worst Case: Continuous \requestmessage-s in \pull)}
  \label{fig:cpu_usage}
  \vspace*{-.2cm}  
\end{figure}

\begin{table}[]
\captionsetup{justification=centering}
\resizebox{0.8\columnwidth}{!}{%
\begin{tabular}{?c?c?cccc?}
    \thickhline
    \rowcolor{yellow!5}
    {\textbf{Time (s)}} & & \multicolumn{4}{c?}{\textbf{$\boldsymbol{\pull}$ (with \requestinterval)}}
    \\ 
    \hhline{~~*4{-}}
    \rowcolor{yellow!5}
    (\anninterval/\geninterval) & \multirow{-2}{*}{\textbf{$\boldsymbol{\push}$}}  & \multicolumn{1}{c|}{\textbf{1 s}} & \multicolumn{1}{c|}{\textbf{5 s}} & \multicolumn{1}{c|}{\textbf{10 s}} & \multicolumn{1}{c?}{\textbf{30 s}}  
    \\ \thickhline
1                              & 19.03\%                        & \multicolumn{1}{c|}{5.58\%} & \multicolumn{1}{c|}{1.95\%}    & \multicolumn{1}{c|}{1.09\%}     & \multicolumn{1}{c?}{0.41\%}     
\\ \hline
2                              & 10.51\%                        & \multicolumn{1}{c|}{3.80\%} & \multicolumn{1}{c|}{1.68\%}     & \multicolumn{1}{c|}{1.00\%}     & \multicolumn{1}{c?}{0.40\%}     
\\ \hline
3                              & 7.26\%                         & \multicolumn{1}{c|}{2.88\%} & \multicolumn{1}{c|}{1.48\%}     & \multicolumn{1}{c|}{0.93\%}     & \multicolumn{1}{c?}{0.39\%}     
\\ \hline
4                              & 5.55\%                         & \multicolumn{1}{c|}{2.33\%} & \multicolumn{1}{c|}{1.32\%}     & \multicolumn{1}{c|}{0.86\%}     & \multicolumn{1}{c?}{0.38\%}     
\\ \hline
5                              & 4.49\%                         & \multicolumn{1}{c|}{1.95\%} & \multicolumn{1}{c|}{1.19\%}     & \multicolumn{1}{c|}{0.81\%}     & \multicolumn{1}{c?}{0.37\%}    
\\ \thickhline
\end{tabular}%
}
\vspace*{.2cm}  
\caption{CPU Usage (\busytime) of \push and \pull in Hotel Scenario\\ (No \requestmessage for \requestinterval in \pull)}
\label{table:eval-runtime1}
\vspace*{-.5cm}  
\end{table}

Similarly, in \pub, $\busytime := \frac{\tsvalue{Res}}{\requestinterval}$,
where \tsvalue{Res} is time to perform \response step and \requestinterval is average time 
between two consecutive \requestmessage-s.
Also, \tsvalue{Res} remains constant, identical to \tsvalue{Ann}. Thus, \busytime is a function of \requestinterval.
In the worst case, when \iotdev continuously receives \requestmessage-s, \requestinterval = \geninterval. 

We consider two types of deployment settings for \pub-enabled \iotdev-s: 
(1) {\it a concert hall}, which is crowded (and expected to have a lot of \requestmessage-s) 
for 8 hours a day, and (2) {\it a hotel}, which is crowded for 16 hours a day.
For a fair evaluation, we simply assume 10 \requestmessage-s per hour to be received by \iotdev 
when the location is not crowded. Figure \ref{fig:cpu_usage} shows the runtime overhead 
comparison between \push and \pull in the worst case, with different scenarios.
In the concert hall scenario, \pub reduces \busytime by $\approx~30$\%,
and in the hotel scenario, it decreases by $\approx~65$\%.

Runtime overhead can be further reduced if \iotdev does not consistently receive 
\requestmessage-s during the crowded time block ($\requestinterval>\geninterval$).
This can occur if users are already informed about nearby devices (no new users are around) or 
if \requestmessage-s are clustered at specific times.
As shown in Table \ref{table:eval-runtime1}, in the hotel setting, \busytime in \pub 
is significantly reduced compared to the \push model.
If \iotdev receives \requestmessage every 10s,
\pub \busytime decreases by at least 70\% compared to \push.

\noindent {\em NOTE1:}
On single-core IoT devices (with TZ-M), only Secure world or Normal world can run at any given time.
Therefore, \pub execution blocks normal functionality of \iotdevsoft\ running in Normal world.
While this is not an issue on multi-core devices, \pub execution still incurs certain
runtime overhead, including the relatively expensive signing operation.

\noindent {\em NOTE2:}
Listening for \requestmessage-s does not interfere with normal device functionality, since the network modem
is typically separate from the primary CPU core(s). However, if \iotdevsoft\ attempts to send any packets while \pub 
is running, \iotdevsoft\ will hang until \pub execution completes. Similarly, if an external entity 
(e.g., a server or a digital twin) attempts to communicate with \iotdevsoft\ while \pub is running, 
its packets might be dropped and would need to be retransmitted.

\noindent {\bf Empirical Evaluation:}
Runtime overhead incurred by \pub execution (and interruption of \iotdevsoft) is measured by comparing 
the runtime of a task with and without \pub.
This experiment is conducted with the sample application, temperature sensor software. 
Furthermore, \pub invocation (and therefore normal operation interruption) depends on variables, 
such as \requestinterval and \geninterval. 
We also measure how changing these parameters, and thus the frequency of 
\pub invocation, impacts the runtime of \iotdevsoft.

For a fair comparison with \paisa, we first evaluate runtime overhead without the lazy-response 
mechanism (i.e., $\geninterval=0$). Recall that \iotdevsoft, a temperature sensor software, 
reads temperature data and sends it to the server every $5$s. We measure \pub overhead added to this task. 
Table \ref{table:eval-runtime-no-tgen} details the runtime overhead for varying \requestinterval-s.

In \pub, if \iotdev receives a \requestmessage every $1$s (i.e., $\requestinterval=1$s), 
\iotdevsoft\ would suffer from a significant delay of $2.13$s ($42.55$\%), on average. 
This is mainly because \pubtcb\ executes $5$ times, on average, during \iotdevsoft's interval of $5$s. 
Each \pubtcb\ execution consumes $\approx~233$ms while stopping \texttt{SysTick}, which is used as
the timer in freeRTOS. In other words, tasks in Normal world are unaware of \pubtcb\ execution in 
Secure world, leading to delayed execution of Normal world tasks.
Unsurprisingly, as \requestinterval grows (i.e., infrequent \requestmessage-s from users), the overhead 
on \iotdevsoft\ sharply decreases.
For example, \iotdevsoft\ overhead is $0.56$s ($11.13$\%) when $\requestinterval=3$s, and $0.22$s ($4.43$\%) 
when $\requestinterval=7$s.

Note that \requestinterval in \pub is equivalent to \anninterval in \paisa when $\geninterval=0$.
For both, \paisa's Annoucnement and \pub's \gen procedures are executed at those intervals.
Both sign and generate the message containing \iotdev device information.
In \paisa, \anninterval is fixed and configured at provisioning time.
In contrast, \requestinterval in \pub depends on the environment.
When there are no new users sending \requestmessage-s, \iotdevsoft\ runs with no interference.
Therefore, \pub incurs lower runtime overhead than \paisa, making it better 
suited for \iotdev-s in less crowded settings.

\begin{table*}[]
\begin{tabular}{?c?c|c|c|c|c|c|c|c|c|c?}
\thickhline
\rowcolor{yellow!5}
\textbf{\anninterval/\requestinterval} & \textbf{1s} & \textbf{2s} & \textbf{3s} & \textbf{4s} & \textbf{5s} & \textbf{6s} & \textbf{7s} & \textbf{8s} & \textbf{9s} & \textbf{10s} \\ \thickhline

\cellcolor{yellow!5}  & 2.13 & 0.89 & 0.56 & 0.40 & 0.32 & 0.26 & 0.22 & 0.20 & 0.17 & 0.15 \\

\cellcolor{yellow!5} \multirow{-2}{*}{\textbf{Mean (s)}} & (42.55\%) & (17.70\%) & (11.13\%) & (8.08\%) & (6.39\%) &	(5.30\%) & (4.43\%) & (3.95\%) & (3.44\%) & (3.08\%) \\ \hline

\cellcolor{yellow!5}\textbf{Std (s)} & 0.10 & 0.06 & 0.11 & 0.14 & 0.07 & 0.10 & 0.14 & 0.15 & 0.15 & 0.15 \\ \hline

\cellcolor{yellow!5}\textbf{Min (s)} & 2.06 & 0.59 & 0.29 & 0.29 & 0.29 & 0.00 & 0.00 & 0.00 & 0.00 & 0.00 \\ \hline

\cellcolor{yellow!5}\textbf{Max (s)} & 2.40 & 0.92 & 0.63 & 0.63 & 0.61 & 0.36 & 0.33 & 0.33 & 0.33 & 0.32 \\ \hline

\cellcolor{yellow!5}\textbf{Median (s)} & 2.09 & 0.90 & 0.59 & 0.30 & 0.30 & 0.30 & 0.29 & 0.29 & 0.29 & 0.29 \\ \thickhline

\end{tabular}
\vspace*{.2cm}  
\caption{\iotdev Runtime Overhead with Varying \anninterval (in \paisa) and \requestinterval (in \dbpaisa)}
\label{table:eval-runtime-no-tgen}
\end{table*}

\begin{table}[]
\resizebox{\columnwidth}{!}{%
\begin{tabular}{?c?cccccccc?}
\thickhline
\rowcolor[HTML]{76A5AF} 
\textbf{$\boldsymbol{\geninterval}$} & \multicolumn{8}{c|}{\cellcolor[HTML]{76A5AF}\textbf{1 s}} \\ \thickhline
\rowcolor{yellow!5}
\textbf{$\boldsymbol{\requestinterval}$} & \multicolumn{1}{c|}{\cellcolor{yellow!5}\textbf{0.1s}} & \multicolumn{1}{c|}{\cellcolor{yellow!5}\textbf{0.3s}} & \multicolumn{1}{c|}{\cellcolor{yellow!5}\textbf{0.5s}} & \multicolumn{1}{c|}{\cellcolor{yellow!5}\textbf{1s}} & \multicolumn{1}{c|}{\cellcolor{yellow!5}\textbf{2s}} & \multicolumn{1}{c|}{\cellcolor{yellow!5}\textbf{3s}} & \multicolumn{1}{c|}{\cellcolor{yellow!5}\textbf{4s}} & \textbf{5s} \\ \hline
\cellcolor{yellow!5} & \multicolumn{1}{c|}{1.43} & \multicolumn{1}{c|}{1.25} & \multicolumn{1}{c|}{1.25} & \multicolumn{1}{c|}{0.88} & \multicolumn{1}{c|}{0.87} & \multicolumn{1}{c|}{0.56} & \multicolumn{1}{c|}{0.40} & 0.32 \\
\cellcolor{yellow!5}\multirow{-2}{*}{\textbf{Mean (s)}} & \multicolumn{1}{c|}{(28.59\%)} & \multicolumn{1}{c|}{(25.04\%)} & \multicolumn{1}{c|}{(24.99\%)} & \multicolumn{1}{c|}{(17.64\%)} & \multicolumn{1}{c|}{(17.47\%)} & \multicolumn{1}{c|}{(11.17\%)} & \multicolumn{1}{c|}{(7.95\%)} & (6.40\%) \\ \hline
\cellcolor{yellow!5}\textbf{Std (s)} & \multicolumn{1}{c|}{0.13} & \multicolumn{1}{c|}{0.12} & \multicolumn{1}{c|}{0.11} & \multicolumn{1}{c|}{0.08} & \multicolumn{1}{c|}{0.08} & \multicolumn{1}{c|}{0.11} & \multicolumn{1}{c|}{0.14} & 0.07 \\ \thickhline
\rowcolor[HTML]{76A5AF} 
\textbf{$\boldsymbol{\geninterval}$} & \multicolumn{8}{c|}{\cellcolor[HTML]{76A5AF}\textbf{2 s}} \\ \thickhline
\rowcolor{yellow!5}
\textbf{$\boldsymbol{\requestinterval}$} & \multicolumn{1}{c|}{\cellcolor{yellow!5}\textbf{0.1s}} & \multicolumn{1}{c|}{\cellcolor{yellow!5}\textbf{0.3s}} & \multicolumn{1}{c|}{\cellcolor{yellow!5}\textbf{0.5s}} & \multicolumn{1}{c|}{\cellcolor{yellow!5}\textbf{1s}} & \multicolumn{1}{c|}{\cellcolor{yellow!5}\textbf{2s}} & \multicolumn{1}{c|}{\cellcolor{yellow!5}\textbf{3s}} & \multicolumn{1}{c|}{\cellcolor{yellow!5}\textbf{4s}} & \textbf{5s} \\ \hline
\cellcolor{yellow!5} & \multicolumn{1}{c|}{0.73} & \multicolumn{1}{c|}{0.71} & \multicolumn{1}{c|}{0.68} & \multicolumn{1}{c|}{0.55} & \multicolumn{1}{c|}{0.40} & \multicolumn{1}{c|}{0.55} & \multicolumn{1}{c|}{0.40} & 0.32 \\ 
\cellcolor{yellow!5}\multirow{-2}{*}{\textbf{Mean (s)}} & \multicolumn{1}{c|}{(14.64\%)} & \multicolumn{1}{c|}{(14.12\%)} & \multicolumn{1}{c|}{(13.53\%)} & \multicolumn{1}{c|}{(10.97\%)} & \multicolumn{1}{c|}{(8.04\%)} & \multicolumn{1}{c|}{(10.95\%)} & \multicolumn{1}{c|}{(8.04\%)} & (6.34\%) \\ \hline
\cellcolor{yellow!5}\textbf{Std (s)} & \multicolumn{1}{c|}{0.13} & \multicolumn{1}{c|}{0.12} & \multicolumn{1}{c|}{0.11} & \multicolumn{1}{c|}{0.08} & \multicolumn{1}{c|}{0.08} & \multicolumn{1}{c|}{0.11} & \multicolumn{1}{c|}{0.14} & 0.07 \\ \thickhline
\rowcolor[HTML]{76A5AF} 
\textbf{$\boldsymbol{\geninterval}$} & \multicolumn{8}{c|}{\cellcolor[HTML]{76A5AF}\textbf{3 s}} \\ \thickhline
\rowcolor{yellow!5}
\textbf{$\boldsymbol{\requestinterval}$} & \multicolumn{1}{c|}{\cellcolor{yellow!5}\textbf{0.1s}} & \multicolumn{1}{c|}{\cellcolor{yellow!5}\textbf{0.3s}} & \multicolumn{1}{c|}{\cellcolor{yellow!5}\textbf{0.5s}} & \multicolumn{1}{c|}{\cellcolor{yellow!5}\textbf{1s}} & \multicolumn{1}{c|}{\cellcolor{yellow!5}\textbf{2s}} & \multicolumn{1}{c|}{\cellcolor{yellow!5}\textbf{3s}} & \multicolumn{1}{c|}{\cellcolor{yellow!5}\textbf{4s}} & \textbf{5s} \\ \hline
\cellcolor{yellow!5} & \multicolumn{1}{c|}{0.48} & \multicolumn{1}{c|}{0.47} & \multicolumn{1}{c|}{0.46} & \multicolumn{1}{c|}{0.40} & \multicolumn{1}{c|}{0.40} & \multicolumn{1}{c|}{0.26} & \multicolumn{1}{c|}{0.41} & 0.32 \\ 
\cellcolor{yellow!5}\multirow{-2}{*}{\textbf{Mean (s)}} & \multicolumn{1}{c|}{(9.65\%)} & \multicolumn{1}{c|}{(9.39\%)} & \multicolumn{1}{c|}{(9.29\%)} & \multicolumn{1}{c|}{(8.06\%)} & \multicolumn{1}{c|}{(8.10\%)} & \multicolumn{1}{c|}{(5.24\%)} & \multicolumn{1}{c|}{(8.20\%)} & (6.37\%) \\ \hline
\cellcolor{yellow!5}\textbf{Std (s)} & \multicolumn{1}{c|}{0.15} & \multicolumn{1}{c|}{0.15} & \multicolumn{1}{c|}{0.15} & \multicolumn{1}{c|}{0.14} & \multicolumn{1}{c|}{0.14} & \multicolumn{1}{c|}{0.10} & \multicolumn{1}{c|}{0.15} & 0.07 \\ \thickhline
\rowcolor[HTML]{76A5AF} 
\textbf{$\boldsymbol{\geninterval}$} & \multicolumn{8}{c|}{\cellcolor[HTML]{76A5AF}\textbf{4 s}} \\ \thickhline
\rowcolor{yellow!5}
\textbf{$\boldsymbol{\requestinterval}$} & \multicolumn{1}{c|}{\cellcolor{yellow!5}\textbf{0.1s}} & \multicolumn{1}{c|}{\cellcolor{yellow!5}\textbf{0.3s}} & \multicolumn{1}{c|}{\cellcolor{yellow!5}\textbf{0.5s}} & \multicolumn{1}{c|}{\cellcolor{yellow!5}\textbf{1s}} & \multicolumn{1}{c|}{\cellcolor{yellow!5}\textbf{2s}} & \multicolumn{1}{c|}{\cellcolor{yellow!5}\textbf{3s}} & \multicolumn{1}{c|}{\cellcolor{yellow!5}\textbf{4s}} & \textbf{5s} \\ \hline
\cellcolor{yellow!5} & \multicolumn{1}{c|}{0.36} & \multicolumn{1}{c|}{0.35} & \multicolumn{1}{c|}{0.35} & \multicolumn{1}{c|}{0.32} & \multicolumn{1}{c|}{0.26} & \multicolumn{1}{c|}{0.26} & \multicolumn{1}{c|}{0.20} & 0.31 \\ 
\cellcolor{yellow!5}\multirow{-2}{*}{\textbf{Mean (s)}} & \multicolumn{1}{c|}{(7.30\%)} & \multicolumn{1}{c|}{(7.06\%)} & \multicolumn{1}{c|}{(7.04\%)} & \multicolumn{1}{c|}{(6.33\%)} & \multicolumn{1}{c|}{(5.22\%)} & \multicolumn{1}{c|}{(5.11\%)} & \multicolumn{1}{c|}{(3.93\%)} & (6.24\%) \\ \hline
\cellcolor{yellow!5}\textbf{Std (s)} & \multicolumn{1}{c|}{0.13} & \multicolumn{1}{c|}{0.12} & \multicolumn{1}{c|}{0.12} & \multicolumn{1}{c|}{0.07} & \multicolumn{1}{c|}{0.10} & \multicolumn{1}{c|}{0.11} & \multicolumn{1}{c|}{0.15} & 0.09 \\ \thickhline
\rowcolor[HTML]{76A5AF} 
\textbf{$\boldsymbol{\geninterval}$} & \multicolumn{8}{c|}{\cellcolor[HTML]{76A5AF}\textbf{5 s}} \\ \thickhline
\rowcolor{yellow!5}
\textbf{$\boldsymbol{\requestinterval}$} & \multicolumn{1}{c|}{\cellcolor{yellow!5}\textbf{0.1s}} & \multicolumn{1}{c|}{\cellcolor{yellow!5}\textbf{0.3s}} & \multicolumn{1}{c|}{\cellcolor{yellow!5}\textbf{0.5s}} & \multicolumn{1}{c|}{\cellcolor{yellow!5}\textbf{1s}} & \multicolumn{1}{c|}{\cellcolor{yellow!5}\textbf{2s}} & \multicolumn{1}{c|}{\cellcolor{yellow!5}\textbf{3s}} & \multicolumn{1}{c|}{\cellcolor{yellow!5}\textbf{4s}} & \textbf{5s} \\ \hline
\cellcolor{yellow!5} & \multicolumn{1}{c|}{0.29} & \multicolumn{1}{c|}{0.29} & \multicolumn{1}{c|}{0.29} & \multicolumn{1}{c|}{0.26} & \multicolumn{1}{c|}{0.26} & \multicolumn{1}{c|}{0.26} & \multicolumn{1}{c|}{0.20} & 0.16 \\ 
\cellcolor{yellow!5}\multirow{-2}{*}{\textbf{Mean (s)}} & \multicolumn{1}{c|}{(5.86\%)} & \multicolumn{1}{c|}{(5.88\%)} & \multicolumn{1}{c|}{(5.89\%)} & \multicolumn{1}{c|}{(5.15\%)} & \multicolumn{1}{c|}{(5.24\%)} & \multicolumn{1}{c|}{(5.26\%)} & \multicolumn{1}{c|}{(3.91\%)} & (3.19\%) \\ \hline
\cellcolor{yellow!5}\textbf{Std (s)} & \multicolumn{1}{c|}{0.05} & \multicolumn{1}{c|}{0.04} & \multicolumn{1}{c|}{0.04} & \multicolumn{1}{c|}{0.11} & \multicolumn{1}{c|}{0.10} & \multicolumn{1}{c|}{0.10} & \multicolumn{1}{c|}{0.14} & 0.15 \\ \thickhline
\end{tabular}
}
\vspace*{.1cm}  
\caption{$\boldsymbol{\iotdev}$ Runtime Overhead with Varying \geninterval in \dbpaisa}
\label{table:eval-runtime-tgen}
\end{table}

Table \ref{table:eval-runtime-tgen} shows \pub overhead with varied \requestinterval-s and \geninterval-s.
As mentioned in Section \ref{sec:implementation}, if there are multiple \requestmessage-s in a given \geninterval, 
\pubtcb\ receives them, appends their nonces to \nlist, and generates a single \responsemessage to respond to all at the end of \geninterval.
For example, when $\geninterval = 1$s, $13$-$15$ \requestmessage-s are collectively handled by one \responsemessage on average with $0.1$ \requestinterval.
This incurs \iotdevsoft\ delay of $1.43$s ($28.59$\%).
Moreover, with the same $\requestinterval = 1$s, \geninterval significantly impacts the overhead:
$2.13$s ($42.55$\%) for $\geninterval=0$s,  $0.88$s ($17.64$\%) for $\geninterval=1$s, and 
$0.26$s ($5.15$\%) for $\geninterval=5$s.

As shown in Table \ref{table:eval-runtime-usr}, a user obtains device information in $3.57$s, on average.
Thus, the result can be shown to the user within $10$s when $\geninterval=5$s.
A too-long \geninterval would lead to users having a false sense of privacy.
It is because they may leave the place before getting \responsemessage-s. Then, they would think that there are no nearby \iotdev-s.
To avoid such issues, \geninterval must be configured reasonably by compliant manufacturers.
This is further discussed in Section \ref{sec:limitation}.

Recall that the network peripheral is configured as secure. \iotdevsoft\ invokes an NSC function to send data 
to the remote server. The overhead of the NSC function call primarily stems from context-switching between 
Secure and Normal worlds. To measure this overhead, the number of cycles was measured before and after calling the NSC function. 
We create a mock function that executes the same task as the 
NSC function in Normal world, and measure the number of cycles to execute the mock function.
The NSC function call requires $519$ cycles, while the mock function takes $392$ cycles.
Thus, the overhead of calling an NSC function is only $127$ cycles, corresponding to $<1\mu$s on 
\iotdev running at $150$MHz.

\begin{table}[]
\captionsetup{justification=centering}
\resizebox{\columnwidth}{!}{%
    \begin{tabular}{?c?cc?cc?cc?}
    \thickhline
    \rowcolor{yellow!5} 
    & \multicolumn{2}{c?}{\bf {NXP (mA)}}         & \multicolumn{2}{c?}{\bf {ESP (mA)}}            & \multicolumn{2}{c?}{\bf \Gape[0pt][2pt]{\makecell{\iotdev (mA) \\ (NXP+ESP)}}} \\
    \hhline{~*6{-}}
    \rowcolor{yellow!5}
      \multirow{-3}{*}{\bf \makecell{Time (s) \\ (\anninterval/\geninterval)}} & \multicolumn{1}{c|}{\push} & {\pull} & \multicolumn{1}{c|}{\push}  & {\pull}   & \multicolumn{1}{c|}{\push}      & {\pull}       \\ \thickhline
    1                                                                                      & \multicolumn{1}{c|}{8.46} & 9.12 & \multicolumn{1}{c|}{60.35} & 100.75 & \multicolumn{1}{c|}{68.81}     & 109.87     \\ \hline
    2                                                                                      & \multicolumn{1}{c|}{8.46} & 8.78 & \multicolumn{1}{c|}{45.43} & 99.13  & \multicolumn{1}{c|}{53.88}     & 107.91     \\ \hline
    3                                                                                      & \multicolumn{1}{c|}{8.45} & 8.67 & \multicolumn{1}{c|}{40.45} & 98.59  & \multicolumn{1}{c|}{48.90}     & 107.26     \\ \hline
    4                                                                                      & \multicolumn{1}{c|}{8.45} & 8.62 & \multicolumn{1}{c|}{37.96} & 98.31  & \multicolumn{1}{c|}{46.42}     & 106.93     \\ \hline
    5                                                                                      & \multicolumn{1}{c|}{8.45} & 8.58 & \multicolumn{1}{c|}{36.47} & 98.15  & \multicolumn{1}{c|}{44.92}     & 106.74     \\ \thickhline
    \end{tabular}
}
\vspace*{.1cm}  
\caption{$\boldsymbol{\iotdev}$ Energy Consumption of \push and \pull}
\label{table:eval-energy}
\vspace*{-.3cm}  
\end{table}

\begin{table*}[]
\captionsetup{justification=centering}
\begin{tabular}{?c?c?c?c?c?c?c?}
\thickhline
\rowcolor{yellow!5} 
\textbf{Model} & \textbf{Event} & \textbf{Time (ms)} & \textbf{Std Dev (ms)} & \textbf{Min (ms)} & \textbf{Max (ms)} & \textbf{Median (ms)} 
\\ \thickhline
\cellcolor{yellow!5} $\boldsymbol{\push}$ & Scan - Display & 1324.56 & 405.81 & 794 & 3085 & 1279 
\\ \thickhline
\cellcolor{yellow!5} & Request - Reception & 2237.18 & 526.17 & 993 & 3721 & 2215.5 
\\ \cline{2-7} 
\cellcolor{yellow!5} & Reception - Display & 1328.60 & 205.99 & 993 & 2244 & 1305.0
\\ \cline{2-7} 
\multirow{-3}{*}{\cellcolor{yellow!5} $\boldsymbol{\pull}$ } & Total & 3565.62 & 519.04 & 2611 & 4952 & 3486.5
\\ \thickhline
\end{tabular}
\vspace*{.2cm}
\caption{$\boldsymbol{\usrdev}$ Runtime Overhead of \push and \pull \\}
\label{table:eval-runtime-usr}
\end{table*}

\subsection{\usrdev Runtime Overhead}\label{subsec:eval-run-usr}
Table \ref{table:eval-runtime-usr} shows the latency of each step in the \pub app when \geninterval $= 0$s.
After sending \requestmessage, it takes $\approx~2.2$s, on average, to receive \responsemessage.
In the worst case, the \pub app would wait $\approx~4.95$s.
This overhead mostly stems from network communication delay between \usrdev and \iotdev.
In both \paisa and \pub, processing \responsemessage takes $\approx~1.3$s, i.e., 
the time between receiving \responsemessage and displaying device information on \usrdev screen.
This latency is primarily due to the fetching of \devmanifest from \devmanifesturl.

In \pub, the app needs to wait for $\approx~3.5$s to receive \responsemessage.
In contrast, in \push model, the app simply listens for device announcements. 
It displays \iotdev details much faster, after $\approx~1.3$s.

\subsection{\iotdev Energy Consumption} \label{subsec:eval-energy}
On the NXP board representing \iotdev, we measure the current by observing the 
voltage drop over a $2.43$ohm resistor via Pinout $12$.
However, the network module (ESP) does not support power consumption measurements.
Hence, our ESP board's current measurement relies on the energy estimation specified in 
the official documentation \cite{esp-power}.
Transmission duration is assumed to be $300$ms with $30$ms intervals for reliable 
reception on \usrdev, i.e., $10$ transmissions per \requestmessage.

Note that the current on the ESP board is substantially higher than
on the NXP board because BLE on the ESP board drains quite a lot of energy:
$97.5$mA for receiving and $130$mA for transmitting, on average. 
Also, the ESP board consumes more energy than mainstream BLE technology 
(Nordic BLE \cite{nordicBLE}) because it is a standalone board running a 
real-time operating system. BLE energy consumption can be lowered by 
integrating BLE into the device.

Table \ref{table:eval-energy} shows energy consumption on NXP and ESP boards 
with \push and \pull models. Power consumption on the NXP board remains almost the same in both.
Since the \pull model continuously listens for \requestmessage-s on the ESP board, it results in constant 
high power consumption. However, energy consumed in the \push model goes down 
significantly when \anninterval increases while making users wait longer to receive 
\responsemessage-s. Consequently, as expected, the \push model is more energy efficient.
Nevertheless, the gap in power consumption of \push and \pull models can be reduced when
(1) BLE is integrated into \iotdev as mentioned above, and
(2) \anninterval is small to minimize the latency to get device information from the user perspective.

The preceding analysis shows a clear trade-off between performance and energy overheads. 
See details in Section \ref{subsec:discussion-tradeoff}.

\begin{figure}[t]
  \centering  
  \captionsetup{justification=centering}  
  \includegraphics[width=\columnwidth]{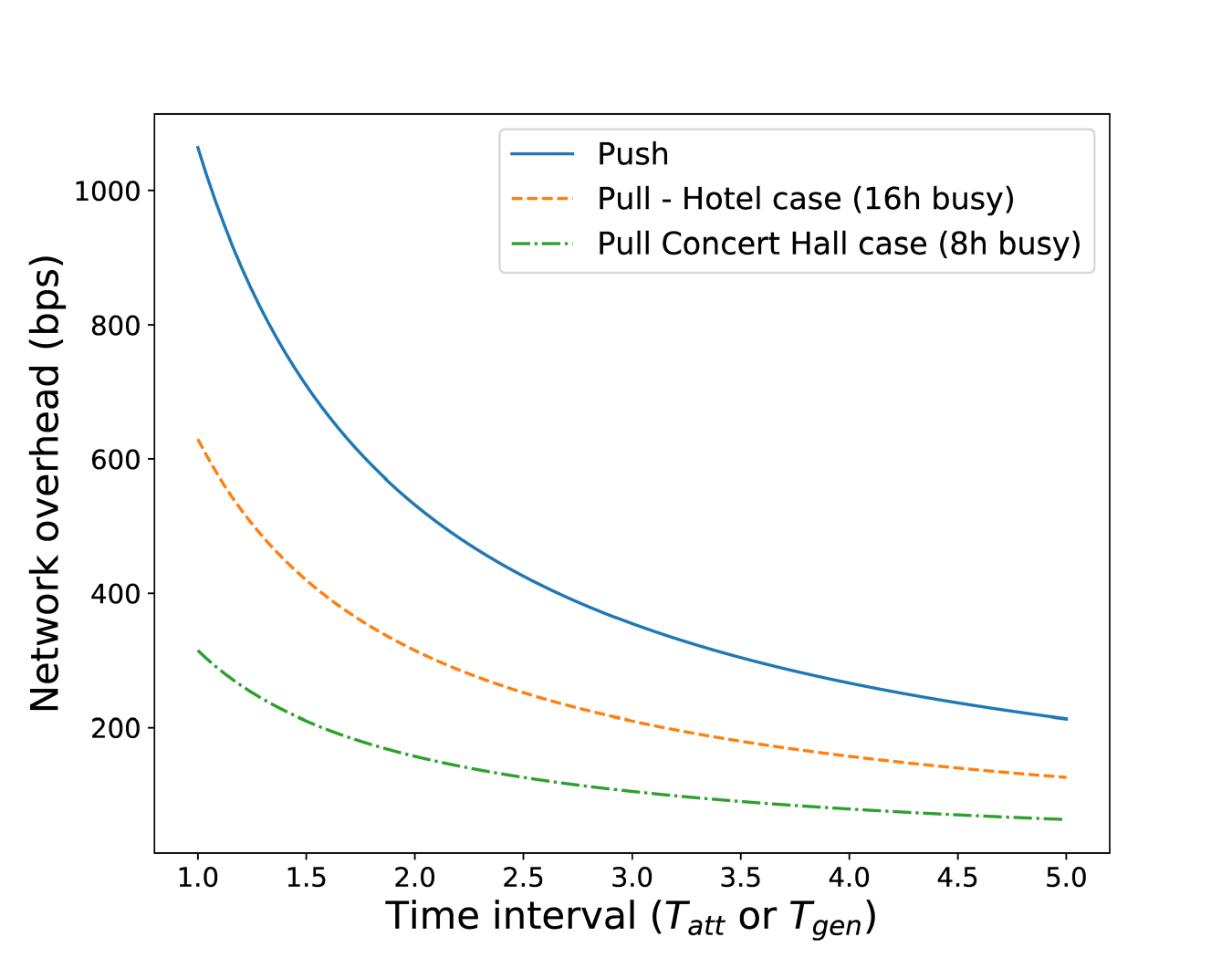}
  \vspace*{-.7cm}  
  \caption{Network Traffic Overhead}
  \label{fig:network_traffic}
  \vspace*{-.3cm}  
\end{figure}

\subsection{Network Overhead} \label{subsec:eval-network-overhead}
In \paisa \push model, announcement size is $128$ bytes, while in the \pull model, \requestmessage and 
\responsemessage are $18$ and $100$ bytes, respectively
Since both \push and \pull models use the same Bluetooth packet header, we compare only the payload size.
Figure \ref{fig:network_traffic} shows bits per second (bps) of \push and \pull \iotdev-s, respectively,
with varying intervals. Bandwidth overhead is reduced by $40.1$\% in the hotel scenario and $70.5$\% in the 
concert hall scenario. However, the \pull model raises another concern; packet size scales with the number of 
\usrdev \requestmessage-s within one \geninterval, leading to larger \responsemessage.

To alleviate this problem, \push and \pull models can be blended to dynamically
switch between \pull and \push protocols depending on the rate of \requestmessage-s.
For instance, given a high influx of \requestmessage-s, \pull model would switch to \push model.
This mitigation is further discussed in Section \ref{subsec:discussion-tradeoff}.

\section{Discussion \& Limitations}\label{sec:limitation}
\subsection{\push\ {\em vs} \pull\ Tradeoffs \& Blending \label{subsec:discussion-tradeoff}}
\push and \pull models have trade-offs in terms of energy consumption as well as performance overheads.
Also, network bandwidth utilization in both models depends on the number of \iotdev-s and \usrdev-s in an area.
The better-suited model is contingent on the deployment use case.
However, it is quite straightforward to blend them and support on-the-fly switching between them.

For example, suppose that \iotdev is using \pull mode. Once it starts receiving an influx of \requestmessage-s (say, over a certain threshold) in rapid succession, it switches to \push mode.
\iotdev stays in \push mode for a certain period (configurable by \mfr) and then switches back to \pull mode.
In other words, when the setting experiences an
influx of new users, \push mode works better, while \pull is better when there are fewer and/or infrequent new users.
Note that \usrdev need not be aware of whether \iotdev is running in \push model or \pull model.
It can collect both announcement messages (from \push devices) and \responsemessage-s (from \pull devices) once it sends a single \requestmessage.
This heuristic is quite easy to implement, in part because announcement messages in \push and
\responsemessage-s in \pull are almost identical format-wise.

\subsection{\requestmessage DoS Mitigation \label{subsec:discussion-dos}}
As discussed in Section \ref{subsec:pull_challenge}, the lazy-response approach does not fully mitigate
DoS attacks: once \nlist size reaches \poolsize, \iotdev composes and signs \responsemessage 
without waiting for \geninterval to elapse. Thus, if it is subject to a high volume of incoming 
\requestmessage-s, \iotdev can be essentially choked. 

Intuitively, public key-based signing can be replaced with highly efficient symmetric MACs. 
However, this would require the existence and involvement of a trusted third party (e.g., \mfr)  
to verify such MACs for \usrdev. (Recall that \iotdev-s and \usrdev-s have no prior security context.)
When \iotdev sends \responsemessage with an HMAC, \usrdev can ask \mfr, via a secure channel, to verify the HMAC.
We consider this approach undesirable as it involves an additional (and trusted) third party, which
can become a bottleneck and would represent a highly attractive attack target.

\subsection{Localization \label{subsec:discussion-localization}}
Localizing \iotdev-s can help considerably improve user awareness and IoT device transparency.
Although \paisa and \pub discover nearby devices, none of them truly and reliably localize them. 

Indoor localization using WiFi or Bluetooth poses a challenge.
Localization typically relies on signal strength (e.g., RSSI), timing information (e.g., time-of-flight, 
time-of-arrival, and round-trip time), or directional information (e.g., angle-of-arrival). 
Signal strength-based localization is relatively imprecise and vulnerable to spoofing attacks 
\cite{spoof1,spoof2, spoof3}. Meanwhile, timing-based localization is vulnerable to distance 
enlargement attacks, whereby the device falsely claims to be farther than it really is \cite{uwbde}.

High-accuracy localization relies on external infrastructure or multiple antennas to triangulate location.
Recently, Bluetooth 5.1 \cite{bluetooth-5.1} introduced a new feature called Direction Finding, which 
supports highly accurate localization.
This requires devices to have multiple antennas to triangulate the locations of other devices.
The emerging Ultra-Wideband (UWB) technology \cite{uwb} is promising for indoor localization 
due to its high precision and resistance to interference. Nonetheless, these advanced 
localization technologies are not yet widely available on commodity IoT devices.

\subsection{TZ-M Alternatives}
\pub relies on TZ-M to provide guaranteed execution.
It can also be deployed on devices with a Root-of-Trust (RoT) that supports a secure timer, secure network peripherals, secure storage, and task prioritization.
As one of the examples of low-end, GAROTA \cite{garota}, an active RoT solution for small embedded devices, supports a secure timer, UART, and GPIO.
It customizes hardware to offer these security features.
Alternatively, RISC-V devices can execute \pub using MultiZone~\cite{multizone}.
It supports Physical Memory Protection (PMP) to isolate secure execution from non-secure tasks.
In addition, I/O PMP (IOPMP) allows configuring peripherals as secure.

\subsection{False Sense of Privacy with Large \geninterval}
Fast response time of \responsemessage through small \geninterval is relevant to users who linger long enough to receive \iotdev\ \responsemessage-s.
Users who run, walk, or cycle by \iotdev's location (i.e., move out of \iotdev's WiFi/BT broadcast range too quickly) may miss \iotdev\ \responsemessage-s.
We expect privacy-concerned users to remain in a given location long enough to receive \responsemessage-s; \geninterval should be configured by manufacturers to be at most a few seconds to retain users' attention.

\section{Related Work}\label{sec:relatedwork}
\noindent {\bf Device discovery:}
In addition to recent work described in Section \ref{subsec:dev-transparency}, some research \cite{das2018personalized,miao2015cloud,privacy-smart-buildings} 
introduces a registration-based approach to enhance the transparency of device presence and capabilities.
The technique provides a scalable registry-based privacy infrastructure, where IoT device owners publish information
about their IoT devices and their capabilities in online registries accessible to other users.
This setup assists users in identifying nearby IoT resources and selectively informs users
about the data privacy practices of these resources. Furthermore, IoT-PPA \cite{das2018personalized} facilitates the discovery of user-configurable 
settings for IoT resources (e.g., opt-in, opt-out, data erasure), enabling privacy assistants to help 
users align their IoT experience with their privacy preferences.

\noindent {\bf Hidden IoT device detection:}
There is also a large body of research focused on discovering hidden devices through either 
emitting/measuring signals with specialized hardware or analyzing network traffic. 

\cite{NLJD,bugdetector,mmwave,iotscan,lapd} employ specialized hardware to detect hidden 
IoT devices using various technologies, including NLJD sensors \cite{NLJD}, millimeter waves 
\cite{mmwave}, software-defined radio (SDR) signals \cite{iotscan}, and time-of-flight 
sensor data \cite{lapd}. While these methods are effective in detecting the presence of 
IoT devices, they are incapable of identifying them.
Moreover, this technique requires users to possess specialized tools to discover IoT devices.

Network traffic analysis involves examining patterns and packets to deduce the presence of devices.
Prior research used this approach to identify devices \cite{devicemien2019,lumos}, detect
active sharing of sensed information \cite{cameraspy}, and localize devices \cite{lumos}. In particular,
\cite{snoopdog} achieves these objectives by establishing causality between patterns in 
observable wireless traffic. However, network analysis is less effective when IoT devices 
communicate infrequently. Also, malicious \sadv can evade detection by not communicating
when users are detected nearby or manipulating its network traffic \cite{iotgan}.

\noindent {\bf IoT privacy:}
Another line of IoT privacy research involves suggesting privacy labels designed for IoT devices that inform users about security and privacy concerns. 
This entails understanding user concerns, incorporating expert recommendations, and offering the privacy 
labels on devices to aid purchasing decisions \cite{iotprivacy-designspace, iotprivacy-label}. 
Another related  research direction explored users' perceptions of risk and their willingness 
to pay for security/privacy features \cite{iotprivacy-attributes,iotprivacy-consumers}.

Automated privacy assistants and consent platforms \cite{iotprivacy-ppa, hong2004architecture, pa3,meng2019towards,utz2023comparing} 
have been proposed to assist users in managing privacy settings for IoT devices that
they encounter, especially, in scenarios where the volume of notifications might overwhelm users.

\section{Conclusion}\label{sec:conclusion}
This work presents \paisaplus, a novel approach to enhance privacy and security in IoT ecosystems through a \pull-based discovery mechanism.
Unlike previous \push-based models, \paisaplus minimizes unnecessary network traffic and interference with device operations by enabling IoT devices to respond to explicit user requests.
Our implementation and evaluation demonstrate \paisaplus's practicality and efficiency, comparing runtime overheads and energy consumption with the \push model.

\begin{acks}
We thank PETS 2025 reviewers for their
constructive feedback. This work was supported in part by funding from 
NSF Award SATC-1956393, NSA Awards H98230-20-1-0345 and H98230-22-1-0308, 
as well as a DARPA subcontract from Peraton Labs.
\end{acks}

\balance

\bibliographystyle{ACM-Reference-Format}
\bibliography{references}

\appendix
\newpage

\section{\paisaplus Variant -- \private \label{sec:inventory}}

\begin{figure}[t]
  \centering
  \captionsetup{justification=centering}
  \includegraphics[width=0.95\columnwidth]{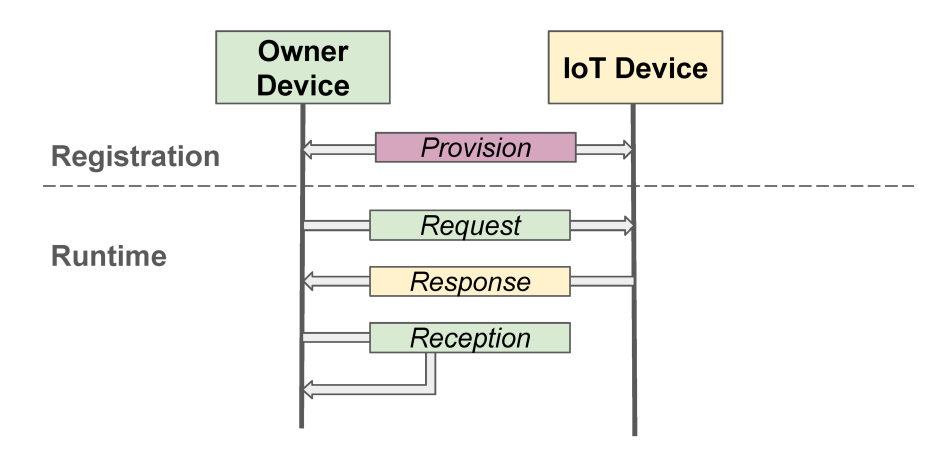}
  \vspace{-0.3cm}
  \caption{\private Overview}
  \label{fig:inventory}
  \vspace{-0.4cm}
\end{figure}

\begin{figure}[t]
  \centering  
  \captionsetup{justification=centering}  
  \includegraphics[width=0.98\columnwidth]{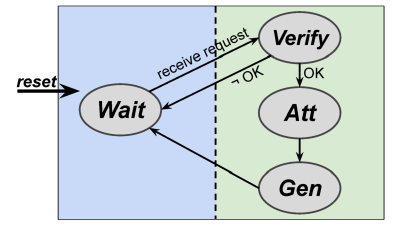}
  \vspace*{-.3cm}  
  \caption{\private State Machine on \iotdev}
  \label{fig:inventory_state_machine}
  \vspace*{-0.35cm}
\end{figure}

\private is a variant of \paisaplus, designed for inventory management in large IIoT settings.
\private has two components: \iotdev and \ownerdev -- the owner's device authorized to 
solicit information from a multitude of \iotdev-s.
As shown in Figure \ref{fig:inventory}, \private also has two phases: {\it Registration} and {\it Runtime}.

\noindent {\bf $\boldsymbol{\registration}$ phase} occurs prior to device deployment.
Each \iotdev is securely provisioned with (1) unique secret key (\seckey) shared with \ownerdev, 
(2) its own device information, and (3) \ownerdev's public key (\ownerpk).

\noindent {\bf $\boldsymbol{\runtime}$ phase} has three steps: \request, \response, and \receptionusr.
There are three significant differences from \pub: (1) \requestmessage is authenticated with 
\ownerdev's private key (\ownersk), (2) \responsemessage is encrypted with \seckey, and
(3) \attest takes place upon every \requestmessage.

\subsection{\private Adversary model} \label{subsec:inventory-adv}
\noindent {\bf DoS attacks on $\boldsymbol{\iotdev}$-s:}
Similar to \pub, malware \sadv can try to deplete \iotdev resources via software vulnerabilities.
However, DoS attacks from a network \sadv are more challenging to address because of costly
verification of \ownerdev's signatures in \requestmessage-s.

\noindent {\bf Replay attacks:}
A network-based \sadv can replay arbitrary messages to both \ownerdev and \iotdev-s.

\noindent {\bf Eavesdropping:}
A network-based \sadv can eavesdrop on all \private protocol messages to learn
individual \iotdev information and the number and types of deployed \iotdev-s.
\sadv can also attempt to link occurrences of one or more \iotdev-s.

\subsection{\private Requirements}
\noindent{\bf System requirements} are the same as in \pub, except that scaling to multiple users is no longer a concern in \private.
Although \private operates in settings with large numbers of \iotdev-s, only an authorized \ownerdev can solicit information from them.

\noindent{\bf Security requirements (beyond those of \pub):}
\begin{compactitem}
    \item {\it Request authentication:} \iotdev must validate each \requestmessage.
    \item {\it Response confidentiality:} \responsemessage must not leak information about \iotdev.
    \item {\it Unlinkability:} Given any two valid \responsemessage-s, the probability of determining if they were produced by the same \iotdev should be negligibly close to 50\%, for any party except \ownerdev.
\end{compactitem}

\subsection{\private Protocol Details} \label{subsec:inventory_detail}
In this section, we only describe the aspects of \private that differ from \pub.

\subsubsection{Registration}
\ownerpk, \seckey, and some metadata are securely installed for each \iotdev. 
However, information, that is not deployment-dependent (e.g., \privatetcb, \iotdevsofthash, and peripheral configuration), is assumed to be securely provisioned earlier by \mfr. 
Note that, since the owner is aware of all identities and types of its \iotdev-s, there is no longer any need for device information to be provisioned on \iotdev or maintained by \mfr.
As a result, \mfr does not play any active role in \private.
The {\bf TCB} of \private is identical to that of \pub.

\subsubsection{Runtime} \label{subsubsec:inventory-runtime}
Similar to \pub, the \private\ \runtime phase has three steps: \request, \response, and \receptionusr.

\paragraph{\bf $\boldsymbol{\privatetcb}$ on $\boldsymbol{\iotdev}$: }
As discussed in Section \ref{subsec:inventory-adv}, DoS attacks by a network-based \sadv are out-of-scope.
Also, scalability is not an issue since only one \ownerdev is assumed.
Furthermore, because \ownerdev's requests are expected to be much less frequent 
than \usrdev's requests in \pub, \attest is performed on the fly upon each \ownerdev's request.

Figure \ref{fig:inventory_state_machine} shows four states of \iotdev:

\noindent{\bf (a) $\boldsymbol{\wait}$:}
The only transition is switching to $\boldsymbol{\verifystep}$ when \requestmessage is received.

\noindent{\bf (b) $\boldsymbol{\verifystep}$:}
\iotdev verifies \requestmessage with \ownerpk.
If verification succeeds, it transitions to $\boldsymbol{\attest}$.
Otherwise, it discards \requestmessage and returns to $\boldsymbol{\wait}$.

\noindent{\bf (c) $\boldsymbol{\attest}$:}
\iotdev computes \attestresult and transitions to $\boldsymbol{\gen}$.
Note that \iotdev need not compute the attestation time or include it in \responsemessage because \attest occurs with every \requestmessage.

\noindent{\bf (d) $\boldsymbol{\gen}$:}
\iotdev encrypts \noncevalue{\owner}{} (\ownerdev's nonce in \requestmessage), device information, and \attestreport with \seckey.
It composes \responsemessage that contains an authentication tag.
Encryption offers \responsemessage confidentiality and unlinkability.

\paragraph{\bf $\boldsymbol{\private}$ app on $\boldsymbol{\ownerdev}$: }
There are two steps on \ownerdev: \request and \receptionusr.

\noindent{\bf \request:} \ownerdev signs \noncevalue{\owner}{} with \ownersk \ and broadcasts \requestmessage, containing \noncevalue{\owner}{} and the signature (\sigvalue{\req}).
After sending \requestmessage, it starts a scan to receive \responsemessage-s from potential nearby \iotdev-s.

\noindent{\bf \receptionusr:}
Upon receiving \responsemessage, \ownerdev retrieves the corresponding \seckey in brute-force attempts. 
After retrieving \seckey, it decrypts \responsemessage with \seckey, and finally, device details are displayed on \ownerdev.

\end{document}